\documentclass[12pt,preprint1]{aastex}
\usepackage{natbib}

\def\spit{{\it Spitzer }}
\def\hip{{\it Hipparcos }}
\def\iras{{\it IRAS }}
\def\iso{{\it ISO }}

\slugcomment{To appear in ApJ, 2007} 

\shorttitle{Dusty Debris Disks Characteristics}
\shortauthors{Rhee et al.}

\begin{document}

\title{Characterization of Dusty Debris Disks: The \iras and \hip Catalogs}

\author{Joseph H. Rhee\altaffilmark{1}, Inseok Song\altaffilmark{1} 
B. Zuckerman\altaffilmark{2}, and Michael McElwain\altaffilmark{2}}

\altaffiltext{1}{Gemini Observatory, 670 North A'ohoku Place, Hilo, HI 
96720; jrhee@gemini.edu, song@gemini.edu}
\altaffiltext{2}{Department of Physics and Astronomy and NASA 
Astrobiology Institute UCLA, Los Angeles, CA 90095; ben@astro.ucla.edu, 
mcelwain@astro.ucla.edu}

\begin{abstract}
Dusty debris disks around main-sequence stars are signposts for the
existence of planetesimals and exoplanets.  From cross-correlating
\hip stars with the \iras catalogs, we identify 146 stars within
120 pc of Earth that show excess emission at 60\,$\micron$.  This
search took special precautions to avoid false positives. Our sample
is reasonably well distributed from late B to early K-type stars, but
it contains very few later type stars. Even though \iras flew more than
20 years ago and many astronomers have cross-correlated its catalogs
with stellar catalogs, we were still able to newly identify debris
disks at as many as 33 main-sequence stars; of these, 32 are within 
100~pc of Earth.  The power of an all-sky survey satellite like \iras 
is evident when comparing our 33 new debris disks with the total of 
only 22 dusty debris disk stars detected first with the more sensitive, 
but pointed, satellite \iso.  Our investigation focuses on the mass, 
dimensions, and evolution of dusty debris disks.

\end{abstract}


\section{Introduction}

Dusty debris disks that surround nearby main-sequence stars were first 
detected by the \textit{Infrared Astronomical Satellite} (\iras) in 1983.  These 
circumstellar disks were inferred from an infrared excess flux between 
25 and 100\,$\micron$ many times brighter than expected from the stellar 
photosphere.  The IR excess was modeled by disk distributions that would 
absorb optical and ultraviolet flux from the host star and then 
isotropically radiate this energy at infrared wavelengths.  The first 
dusty debris disk was discovered around the bright main-sequence star 
Vega \citep{aum84}; consequently a dusty disk around a main-sequence 
star is commonly referred to as the Vega phenomenon.

Numerous studies of T Tauri stars dating back many years indicate the 
characteristic timescale for the dispersal of a surrounding dusty, 
gaseous disk is a few million years.  Following dissipation of the gaseous 
component, the remaining dust can further dissipate during the following 
few million years via coagulation into large objects,
Poynting-Robertson and stellar wind drag, radiation pressure, and
collisional destruction (e.g., \citealt{bac93,lag00,dom03,pla05}).
Vega-like stars are, however, generally much older than 10 Myr, thus the
observed dust should be of secondary origin, most likely replenished via
collision and fragmentation of planetesimals.  Furthermore, the
Vega-phenomenon overlaps with the important planetary system formation
epochs in our solar system: giant gas planet formation within $\sim$10 
Myr, terrestrial planet formation within $\sim$30 Myr, and the era of heavy 
bombardment in the inner solar system within $\sim$600 Myr.  Therefore, 
studies of IR-excess stars can provide crucial information on extrasolar 
planetary formation and evolution.

During the past two decades, about two dozen papers have been published 
that describe \iras, Infrared Space Observatory (\iso), and \spit \textit{Space 
Telescope} searches for stars with excess IR emission ($\S$~3; 
\citealt{lag00,zuc01,dec03}).   These searches employed different 
techniques for cross-correlating IR and stellar sources with no consistent 
definition of an IR excess.  To date, several hundred main-sequence 
IR-excess stars have been reported in the literature including 
those that have an IR excess at 25\,$\micron$.  

A major goal of debris disk research has been to characterize the
temporal evolution of the quantity of dust present in the disks.
Notwithstanding almost two decades of debris disk research using
data from three IR satellites, a convincing assessment of this
temporal evolution remains incomplete.  Such an assessment requires a
large sample of stars and a reliable estimate of the dust mass and the
age for each debris disk system.  False positive IR-excess stars due 
to the large beam size of \iras and improper search or calibration 
techniques have contaminated some previous studies.  Such contamination 
of the debris disk population has not only plagued many follow-up 
observations from ground and/or space observatories but also precludes 
a global assessment of the distribution and evolution of the dust 
population.  

If the IR excess 
is from a bona fide dust disk, then the best estimator of dust mass
comes from submillimeter flux.  Unfortunately, submillimeter flux
measurements are difficult, time-intensive observations.  A more
readily accessible observable is $\tau$, the ratio of excess infrared
luminosity due to dust divided by the total energy output from a star.
We compute values of $\tau$ for each of the IR-excess stars presented
in this paper.  In $\S$~5.1 we discuss the relationship between
submillimeter flux and $\tau$ for those Vega-like stars for which both are
known, and we derive our own relationship, which is used to predict a
dust mass if both $\tau$ and the dust disk radius are known.

Furthermore, estimation of stellar age is often troublesome since most
nearby IR-excess stars are isolated field stars.  In order to obviate
the shortcoming of stellar age estimation, several groups are using
\spit to search for IR-excess stars in nearby young stellar groups
with well-determined ages (e.g., \citealt{sta05}).  However, because
the distance to all rich clusters is substantial (except for the
Hyades), it remains difficult to obtain statistically significant
results even with \spit \citep{sta05}.  Thus, the large, clean sample
of relatively nearby field stars we discuss in the present paper can
contribute in a statistically meaningful way to our understanding of
the Vega phenomenon and its evolution with time.

\section{Search Criteria and Selection Technique}

Zuckerman \& Song (2004a, hereafter Paper I), relying primarily on data in
Silverstone's (2000) thesis, analyzed 58 strong IR-excess stars following
careful checks against possible contamination from various
sources.  Zuckerman \& Song argued that the Vega-like stars are
signposts for the existence of planets and focused their efforts on
identifying stars that would make the best targets for adaptive optics
and precision radial velocity searches.  The present paper extends the
sample analyzed in Paper I in a couple of ways. First, we
significantly increase the sample size so that it is now possible to
address circumstellar dusty disk evolution in a statistically
meaningful way.  This increase is achieved by systematically
cross-correlating all \hip main-sequence stars with 60\,$\micron$
\iras sources in the Faint Source (FSC) and Point Source (PSC)
catalogs.  Our distance limit is 120 pc compared with the 100 pc
adopted in Paper I. Second, the spectral energy distribution (SED)
fitting routine was enhanced with the employment of filter response
functions and a fully automated fit with a $\chi^{2}$ minimization
method.

\begin{figure}
\plotone{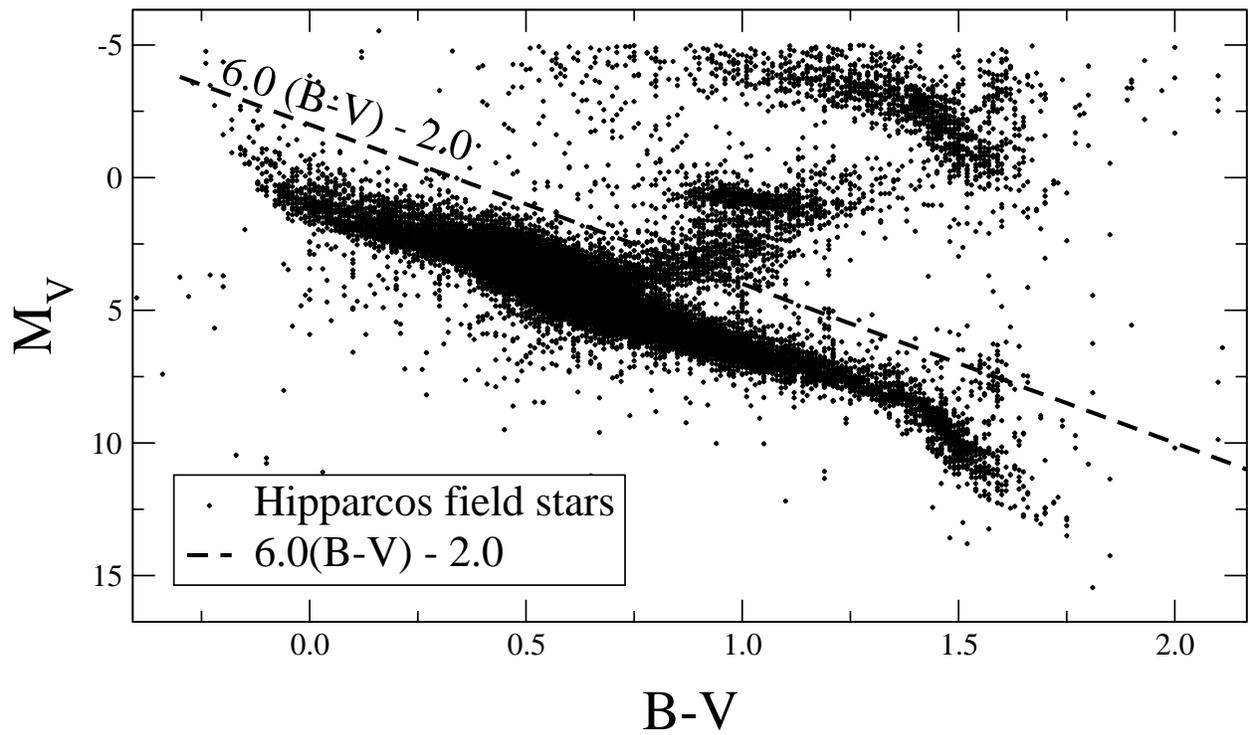}
\caption{Hertzsprung-Russell diagram of \hip field stars.  Stars below the dashed 
line and with \bv $>$ -0.15 have been searched for \iras 60\,$\micron$ excess emission.}
\label{hipMS}
\end{figure}

\hip and \iras data were cross-correlated to search for IR-excess 
stars.  Many sources in the FSC and PSC with optical stellar
identifications are, however, giant stars \citep{ode86,zuc95a}.  A
constraint on the absolute visual magnitude $M_{V}$ $\geqq$ 6.0(\bv) -
2.0 (Fig. \ref{hipMS}) was applied to the entire 118,218 stars of 
the \hip catalog to remove giant stars from our sample.  This 
cut eliminated 50,164 stars, leaving 68,054 \hip stars for further 
investigation.

These pre-screened \hip dwarfs were then cross-correlated
against \iras sources.  The \iras FSC was used to cross-correlate the
53,157 stars located out of the Galactic plane ($|b|$ $>$
$10^{\circ}$), while the PSC was used for the 14,897 stars in the
Galactic plane and to recover any object missed by the FSC out of the 
Galactic plane.  All FSC sources with a detection at 60\,$\micron$ (i.e. 
a 60\,$\micron$ flux quality of 2 or 3) and a \hip dwarf within 
45$\arcsec$ were selected for further investigation.  A search radius 
of 45$\arcsec$ was adopted to reflect the average FSC 3~$\sigma$ 
positional error.  For PSC sources in the Galactic plane, \hip 
dwarfs within only a 10$\arcsec$ search radius were retained, in 
order to avoid contamination of spurious sources in the
crowded fields of the Galactic plane.  There were 557 stars (481 from
the FSC and 76 from the PSC) that passed the initial
cross-correlation.  Unfortunately, the FSC is only $\sim$80\%
complete.  We therefore cross-correlated all main-sequence \hip
stars outside the Galactic plane with the PSC using a search radius 
of 45$\arcsec$.  We found an additional
65 stars in the PSC that had 60\,$\micron$ detections, but were
unidentified in the FSC.  Most of these stars from the PSC were
detected at 12\,$\micron$ (but not 60\,$\micron$) in the FSC.  In
contrast, \citet{sil00} cross-correlated \hip and \iras
FSC sources only.  Our correlation with the \iras FSC and PSC left a
collection of 622 main-sequence stars identified in the \hip
catalog that had 60\,$\micron$ counterparts detected by \iras.

In young and massive main-sequence stars, significant IR flux arises 
from free-free emission.  Such stars, namely spectral types O1-B5, 
were excluded from our sample by rejection of objects with $B-V$ 
$<$ -0.15.  Then a distance cut of 120 pc was applied 
to our sample to avoid contaminations arising from star-forming regions 
and interstellar cirrus as described, for example, in \citet{kal02} 
(see below).

A visual inspection of the remaining excess candidates for the
presence of a background galaxy was conducted by correlating the FSC
and PSC catalogs with NASA's Extragalactic Database (NED) in Digital
Sky Survey (DSS) images.  Any star with a noticeable galaxy within the
3$\sigma$ \iras positional error ellipse was removed from our sample,
and any star with a bright star within the 3$\sigma$ error ellipse was
flagged for further checking of its SED.  Since NED is not complete,
we also carefully checked any DSS optical extended sources (mainly
galaxies) that were not included in NED.  Using the long format of
the FSC catalog, \citet{sil00} compared the 60\,$\micron$ position to
the stellar position and excluded stars whose 60\,$\micron$ offsets are
$>$30\arcsec.  Instead of imposing such a strict constraint on our
sample, we exclude stars only if their 60\,$\micron$ offsets are
greater than the 3 $\sigma$ \iras positional error.  For all FSC
sources, we carefully checked their 60\,$\micron$ positions against any
galaxy or bright nearby star.

For stars with apparent detections in the \iras 100\,$\micron$ band, we
tested for possible contamination from interstellar cirrus.  Some
relatively distant, previously known, IR-excess candidates are
contaminated by interstellar cirrus \citep{kal02}.  We checked the
\iras cirrus flag of all 100\,$\micron$ sources and rejected those with
cirrus flag $>$ 3 except HIP 77542.  HIP 77542 had significant excess
at all wavelengths and was fit nicely with a single blackbody
temperature (Paper I).

A fully automated SED fitting technique
using a theoretical atmospheric model \citep{hau99} was used to
predict stellar photospheric fluxes.  This fit technique is unlike
previous excess searches that use the ``empirical'' color of main
sequence stars to estimate stellar photospheric fluxes.  For each
star, fluxes at $B$, $V$, $J$, $H$, and $K_{s}$ were employed to fit 
the model spectra of a stellar photosphere.  The standard Johnson $B$ 
and $V$ magnitudes were obtained by converting Tycho $B$ and $V$ 
magnitudes using Table~2 in \citet{bes00}.  For the 10 \hip 
objects that did not have observed Tycho $B$ and $V$ magnitudes, 
$B$ and $V$ values were obtained from SIMBAD.  Observed $J$, $H$,
and $K_{s}$ magnitudes came from the 2MASS catalog.  When any star was
brighter than 5th magnitude at $J$, $H$, or $K_{s}$ in 2MASS, we set
its uncertainty to 0.400 mag.  The zero magnitudes in \citet{cox00}
were used to convert the observed magnitudes into a flux density (janskys).
The current Hauschildt et al.\ stellar photosphere model (T. Barman 2004,
private communication) is available for effective temperatures from
1700 to 10,000~K (in 100~K increments from 1700 to 3000~K and in
200~K increments from 3000 to 10,000~K).  The stellar radius and
effective temperature were used as free parameters to fit the observed
fluxes with a $\chi^{2}$ minimization method.

We created model fluxes at each band by convolving each filter
function with the model spectra.  This method provides a more 
accurate representation of the observed flux especially where the 
passband includes significant spectral features such as the Balmer jump.  
Comparing the best-fit model spectra with the observed fluxes, we found 
that the model spectra always overestimated the $B$- and $V$-band fluxes.  
This perhaps arises from some missing opacity sources in the $B$- and 
$V$-bands of the model spectra.  For consistency, we manually set the 
uncertainties of $B$ and $V$-band magnitudes
to 0.25 if the given uncertainty value is smaller than 0.25 mag to
ensure a better fit.

Once the stellar photosphere was modeled, a dust component was fit 
with a blackbody curve.  \iras upper limits were not included in the 
dust fitting, but we mandated that the upper limits are always above 
the estimated total (star and dust) flux.  Temperature dependent 
\iras color corrections should be carefully considered.  Both the 
stellar photosphere and dust emission contribute to the observed 
\iras flux as follows;

\begin{equation}
F^{obs}_{\iras} = F^{unc}_{phot} + F^{unc}_{dust}
\end{equation}

where the subscript ``unc'' stands for ``uncorrected.''
Thus, accurate estimation of a color correction value requires not only 
the flux of the stellar photosphere but also that of the dust, which is 
obtained through the blackbody fitting.  But the problem is that 
both dust flux and the color correction are a function of dust 
temperature, which requires an iterative process to determine the 
dust temperature in color-corrected \iras dust flux.  Instead we 
obtained the dust temperature by fitting the uncorrected \iras fluxes.  
First, we ``colored'' the stellar photosphere (eq. [2]) 
by multiplying the appropriate color correction terms (${K}_{star}$) 
before subtracting the stellar photosphere  (eq. [3]):

\begin{equation}
F^{unc}_{photosphere} = F_{phot,m} \times K_{star}
\end{equation}
\begin{equation}
F^{unc}_{dust,m} = F^{obs}_{\iras} - F_{phot,m} \times K_{star}
\end{equation}

, where the subscript $m$ stands for ``model.''  
Then we fit the remaining \iras fluxes with the ``colored'' blackbody
curve (eq. [4]),

\begin{equation}
F^{unc}_{dust,m} = F_{dust,m} \times K_{dust}
\end{equation}

By combining equations (3) and (4) and using the stellar photosphere 
model described above, we obtained the best-fit temperatures of the 
stellar and dust emission.  Then the correct total \iras color correction 
terms were calculated by estimating the fractional color terms using 
a weighted average of photosphere and dust fluxes at each wavelength 
(eq. [5]):

\begin{equation}
K_{total} = C_{1} \times K^{bestfit}_{star} + C_{2} \times K^{bestfit}_{dust}
\end{equation}
where C$_{1}$ and C$_{2}$ are the fractional contributions of the stellar 
photosphere and dust to the total measured flux,
\begin{equation}
F^{true}_{\iras} = F^{obs}_{\iras} / K_{tot}
\end{equation}

In displaying the \iras observed magnitudes, we applied the prorated color
correction terms to the \iras measurements (eq. [6]).  As in photosphere
fitting, we created synthetic fluxes at each \iras band by convolving \iras
filter functions with the blackbody curve.

SED fits were performed for all identified \iras and \hip stars,
yielding very precise estimation of stellar photospheric fluxes
(Fig. Set 2).  When available, additional fluxes from
\iso\footnote{\iso measurements were taken from \citet{sil00} and
\citet{hab01}.} and/or \textit{Spitzer}\footnote{These \spit MIPS
measurements were taken from the references given in $\S$~3.}
measurements were used to better fit dust components.  Four objects
were dropped from our list due to possible cirrus contamination or
no 60\,$\micron$ excess based on \iso measurements reported in
\citet{sil00} except HIP 111278.  For some objects, \spit Multiband 
Imaging Photometer for \spit (MIPS) data are available in the
public archive but were not yet published.  In such cases, we extracted
photometry from MIPS pipeline data at 24 and 70\,$\micron$.  No
photometry was attempted on MIPS 160\,$\micron$ pipeline data because
of heavy contamination from a known ``blue leak.''

Several Class I and II pre-main-sequence (PMS) stars were found from 
our SED fits, in which a typical SED of a Class I/II PMS star shows 
large $B$ and $V$ fluxes above the model spectrum and strong but 
flat excess in the IR.  Because we are searching for IR excess among 
main-sequence stars, Class I and II PMS stars were subsequently 
eliminated from our sample.  For completeness, the \iras-identified 
\hip class I \& II PMS stars within 120 pc are listed in Table~1.

Many sources that passed the visual check, especially nearby stars,
showed no IR excess in their SED.  Color corrected \iras fluxes were
compared to the estimated photospheric fluxes.  Stars with no IR
excess ([F$_{\iras}$ - F$_{phot}$] / $\sigma_{\iras}$ $<$ 2.5)
were eliminated except HIP~71284, where $\sigma_{\iras}$ is the \iras 
60\,$\micron$ flux density uncertainty.  IR excess at HIP~71284 was
confirmed by an \iso observation (Paper I).  One hundred forty-six stars 
had IR excess ([F$_{\iras}$ - F$_{phot}$] / $\sigma_{\iras}$ $>$ 3.0), 
and nine stars showed marginal IR excess (2.5 $<$ [F$_{\iras}$ -
F$_{phot}$]/ $\sigma_{\iras}$ $<$ 3.0).  These marginal IR 
excess stars fall into a statistical domain in which $\sim$0.5\% of 
non-excess stars may produce a false excess assuming Gaussian noise 
under \textit{pure statistical detection errors}.  Recent \spit 
observations show that three stars (HIP~65109, HIP~105090, \& HIP~105858) 
that had marginal IR excess from \iras are not IR-excess stars.  In 
addition, even some stars with [F$_{\iras}$ - F$_{phot}$] / 
$\sigma_{\iras}$ $>$3.0 turn out to be false positives.  For example, 
HIP~83137, passing all the tests above, had [F$_{\iras}$ - 
F$_{phot}$] / $\sigma_{\iras}$ = 4.3 and was considered one of 
the better new IR-excess candidates.  However, recent \spit MIPS 
observations found no excess emission at 70\,${\micron}$ at HIP~83137 
along with six other similar stars (HIP~8102, HIP~42913, HIP~49641, 
HIP~75118, HIP~98025, \& HIP~104206).

All six bogus excess stars had \iras excess emission detected at
60\,$\micron$ only.  To date, \spit has looked at a total of 26 such
stars in our sample producing a false excess rate of 27\% (7/26).
Applying this rate to the remaining 54 stars with infrared excess
emission detected at \iras 60\,$\micron$ alone, we anticipate that 
about 15 objects or 10\% (15/146\footnote{146 = 146 + 9 (with marginal 
IR excess) + HIP~71284 - 10 bogus stars.}) of our sample may turn out 
as non-excess stars.

Generally, a bogus excess can be produced in two quite different ways.  
One way is where a real, background, far-IR source is present in the beam 
when \iras pointed toward a \hip star.  The other is where a 3\,$\sigma$ 
noise bump happens to fall near a \hip star.  Apparent excess sources 
rejected for both classes of reasons are listed in Table A4.  The number of 
real background sources (mostly galaxies) anticipated in our sample can be 
estimated in a way analogous to that described in Section 2 of \citet{zuc95a}; 
such an estimate agrees reasonably with the number of 
background galaxies listed in Table A4.

If the background noise has a normal distribution, then we anticipate 
that about one star in 500 could be contaminated by a 3.1\,$\sigma$ noise 
fluctuation.  After our distance and color cuts described above, we were 
left with $\sim$25,800 \hip dwarfs.  Thus, of these, $\sim$50 might be 
contaminated by a noise fluctuation.  Some constraint is supplied by 
examination of \iras SCANPI traces which sometimes show the 60\,$\micron$ 
peak position to be displaced from the stellar position.  Background noise 
could be responsible, in total, for $\sim$20 \hip stars listed in 
Table~A4. 

Nearby M-type stars are now known not to be strong IR-excess sources (e.g. 
\citealt{pla05,ria06}), indeed the only one listed in Table 2 is 
AU~Mic which is a very young star.  There are $\sim$900 
M-type dwarf stars in the \hip catalog and the only one other than AU 
Mic that appeared in our cross correlation with \iras was AX~Mic, in which, 
however, a \spit MIPS observation showed that there is no 70\,$\micron$ 
excess.  According to the above estimates, we might have expected two 
bogus \iras associations in these 900 stars, in reasonable agreement with 
the one, AX~Mic that was actually found.  

We finally present 146 \iras identified \hip IR-excess dwarfs in this
paper.  Among them 33 stars are newly identified as IR-excess stars from
our survey, and only two objects out of these 33 newly identified IR-excess
stars have marginal IR-excess (2.5 $<$ [F$_{\iras}$ - F$_{phot}$] /
$\sigma_{\iras}$ $<$ 3.0).


\section{Overview of previous \iras, \iso, and \spit surveys for dusty debris
disks}

Comparison of \iras with \iso and \spit demonstrates the power of all-sky
surveys.  Notwithstanding that \iras flew more than 20 years ago, through
careful analysis of its database, we have been able to discover perhaps as
many as 33 main-sequence \hip stars with previously unrecognized
dusty debris disks detected at 60\,$\micron$ wavelength. In comparison,
only 22 new 60\,$\micron$ excess stars were discovered in all \iso programs
while $\sim$20 new 70\,$\micron$ excess stars were announced in the 2004
and 2005 \textit{Spitzer}-based literature (see below for references).
Although \iso and \spit have higher sensitivities than \iras, they are both
pointed satellites with a much smaller sky coverage.

\iras surveys and, significantly, some of their limitations are
summarized in $\S$~1 of the present paper and in $\S$~3 of
\citet{zuc01}.  Previous to the present study, \citet{sil00} represented 
the most comprehensive search of the
\iras catalogs for Vega-like 60\,$\micron$ excess stars. However,
Silverstone's primary goal was to use \iso to detect dust at F- and
G-type stars inconclusively detected by \iras at 60\,$\micron$.  He did
not analyze his \iras findings, and his search never reached
publication.  Thus, no \iras survey published prior to 2005 is germane
to issues addressed in the present paper.

\iso was a pointed satellite of modest sensitivity, and surveys by
various groups added relatively few new Vega-like stars.
\citet{dec03} give a comprehensive account of these surveys, a major
goal of which was characterization of the time dependence of the Vega
phenomenon.  One limitation of these studies, as noted by Decin et al.,
is the quite uncertain ages of many of the excess stars.  Indeed, we
disagree with some of the ages in Table~1 of Decin et al.  They
describe some limitations to the results presented by \citet{spa01},
limitations due, in part, to the poorer than expected sensitivity of
\iso.

A next advance was by \citet{man05}, who focused on deducing the
lifetimes and temporal evolution of the dust around the Vega-like stars.
In an innovative analysis, they considered the relative sky-plane velocity
dispersions of the Vega-like stars and of \hip stars in general to
demonstrate that, at any given spectral type, the Vega-like stars are, on
average, younger than the general population of field stars.  They also
showed that the average $\tau$ of the Vega-like stars declines with
increasing velocity dispersion, that is, with increasing age.  Because
their analysis technique is very different from our's and because their
sample of excess stars is not called out explicitly in their paper, it is
not possible to make a direct comparison between their results and ours.
However, wherever their conclusions and ours do overlap, they appear to
be consistent.

Most recently, \citet{moo06} compiled a list of 60 debris disks with
high fractional dust luminosity, $\tau > 10^{-4}$, and within 120 pc
of Earth by searching the \iras and \iso database.  Forty-eight objects in Moor
et al.\ are included in our survey, while 12 objects are absent.  Among
those 12 objects missing, four are not \hip stars, and six of eight 
\hip stars did not have a detection at 60\,$\micron$ with \iras and, 
therefore, did not satisfy our search criteria ($\S$~2).  The remaining
two, HD~121812 (HIP~68160) and HD~122106 (HIP~68380), are rejected in
the present paper due to possible cirrus contamination and the presence of a
nearby galaxy, respectively (see Table~A4 for the list of rejected
sources).  We included five objects (HIP~13005, HIP~25790, HIP~69682,
HIP~77163, and HIP~83480) from the Moor et al. list of rejected
suspicious objects; our reasoning is discussed in the notes for these
individual objects in Table~2.

Five papers that appeared in 2004 or 2005 report \spit detections at
70\,$\micron$ for a total of $\sim$20 Vega-like stars that had not
previously been detected at 60\,$\micron$ by \iras and/or \iso
\citep{mey04,che05,bei05,low05,kim05}. Although it is not possible to
tell exactly how many stars \spit pointed toward (searched) at 70
$\micron$ in these studies, it appears to be of order a few hundred.
Thus, only about 10\% of stars reveal far-IR dust emission at levels
between \iras and \spit sensitivities.

\section{Sample Characteristics}

Our IR-excess sample consists of 146 \hip dwarfs within 120 
pc of Earth.  Figure \ref{dist_bv} illustrates the distance and
$B-V$ distribution of the sample.  The relative paucity 
of debris disks from late-type stars has been previously well 
established and attributed to the \iras detection threshold \citep{son02b}.  
However, grain removal by stellar wind drag at M-type stars could also 
be implicated \citep{pla05}.

Our stars are listed in Table~2 including 51 out of 58 stars from
Paper I.  The remaining seven objects had \iso detections but lacked an
\iras 60\,$\micron$ detection, an absolute requirement in the present
paper.  The \hip and the HD numbers are listed in columns (1) and
(2), respectively.  Spectral type, $V$ magnitude, and distance from Earth from the
\hip main catalog are given in columns (3), (4) and (5), respectively.  The
stellar radius and temperature, $R_{\star}$ (col. [6]) and $T_{\star}$
(col. [7]) are obtained from the SED fit.  As described in $\S$ 2, 
the fitting process was improved from the version
used in Paper I and for some objects the best fit $R_{\star}$ and
$T_{\star}$ deviate slightly from Paper I.  For example, HIP 42430 was
fit with $R_{\star}$ of 1.83 $R_\odot$ and $T_{\star}$ of 5600 K in
Paper I, but the improved fit gives $R_{\star}$ of 1.73 $R_\odot$ and
$T_{\star}$ of 5800 K in Table~2.  Our estimations of $R_{\star}$ are in
good agreement with direct measurements such as those with the
Very Large Telescope (VLT) interferometer as illustrated in Paper I.
The accuracy of our stellar radius measurements is discussed in more
detail in a separate paper (S. Kim et al. 2007, in preparation).

A single-temperature blackbody fit to the dust component yields
$T_{dust}$ (col. [8]) for each star, assuming blackbody radiation from
dust grains in an optically thin disk.  In the case of an \iras
detection at 60\,$\micron$, but with only upper limits at 25 
and 100\,$\micron$, we set $T_{dust}$ at 85 K so that the combined flux
of the star and dust peaks near 60\,$\micron$.  This approach leads to
a conservative estimate of $\tau$ (col. [11], (= $L_{IR}/L_{bol}$).
Additional measurements from \spit and/or \iso were used to better
constrain dust temperature for stars in which such values are
available in the literature or from our calculations. (see $\S$~2).

The characteristic orbital semimajor axis of dust particles, $R_{dust}$, 
is derived from $R_{dust}$ = ($R_{\star}$/2)($T_{\star}$/$T_{dust})^{2}$ 
and listed in column (9) in AU.  The corresponding angular separation 
(arcseconds) between dust particles and the star is indicated in column 
(10).  The conservative nature of $R_{dust}$ and the angular separation 
-- in the sense that the actual value of $R_{dust}$ at a given star 
may be substantially larger than the value given in column (9) -- is 
discussed in detail in Paper I.  Using a simple model of a thin dust 
ring (see $\S$~5.1), dust mass (col. [12]) was estimated for 61 stars 
whose dust excess was detected at two or more wavelengths and whose 
dust radii lie between 9 and 100 AU.  Table~2 lists dust 
mass for a total of 78 stars including 17 stars for which dust mass 
was obtained directly from submillimeter measurements.

Estimation of the age of a star that belongs to a known kinematic
stellar group \citep{zs04b} is relatively straightforward.  For 
stars not presently known to be a member of such a group, age 
estimation is quite difficult and requires cross-checking of 
several different techniques (\citealt{dec03,zs04b} and references 
therein).  The age estimate and age estimation methods for each 
star are given in columns (13) and (14), respectively.  We follow the same 
lettering convention for each method as indicated in Paper I.  A 
comprehensive review of different techniques of age estimation is 
found in \citet{zs04b}.  

When available, confirmation of dust excess from MIPS and/or 
\iso measurements are indicated in column (15) and additional notes 
for individual objects are marked in column (16).  For completeness,
we repeat the notes of Table~1 from Paper I in this paper.  Finally, a 
list of rejected sources and the reason for rejection from our survey 
are presented separately in Table~4.

\section{Dust Evolution over Time}

Figure \ref{age} illustrates the temporal evolution of $\tau$.  The
spectral type of each star is represented by the color of each circle,
from dark blue for B-type to red for M-type.  Circle size reflects the
quality of our estimate of age; large, medium, and small circles depict
good-, normal-, and low-quality age estimates, respectively, as given in
column (13) of Table~2.  The following list summarizes some
characteristics indicated by the distribution of stars in Figure \ref{age}.

\begin{enumerate}
\item\label{item:1} For stars with ages between $\sim$10 Myr and 1 Gyr,
the mean $\tau$ of stars with detectable excess emission declines in 
proportion to (age)$^{0.7}$, but with a dispersion in detected $\tau$ 
of a factor $\sim$30 at a given age.
\item\label{item:2} The percentage of nearby stars with 60\,$\micron$
excess emission detectable by \iras diminishes with increasing stellar
age.
\item\label{item:3} The minimum detected $\tau$ is $\sim$$10^{-5}$ for 
early-type (B, A, and F) stars and $\sim$$10^{-4}$ for later types.  This is 
due to \iras sensitivity limits and the uncertainty of photospheric flux 
estimation.
\item\label{item:4} At any given age, late-type stars tend to have the
largest $\tau$.

\end{enumerate}

As we mentioned in $\S$~3, no pre-2005 analysis of \iras data is germane
to the time evolution of fractional dust excess, $\tau$.  By contrast,
three teams \citep{hab01,spa01,dec03} investigated the temporal
evolution of the dust using the \iso database.  All three studies suffer 
to some degree from small numbers of detected \iso sources or 
uncertain/incorrect stellar ages or both.  \citet{dec03} noticed that 
there are few young stars with $\tau < 10^{-4}$, which also appears in 
our Figure 4.  This rarity of young low $\tau$ stars may be due to the 
fact that there are not many young early-type stars in the solar vicinity 
(say $\lesssim$ 50 pc).

We can roughly quantify item (2) by dividing the \iras stars into three 
age bins, (a) 10-50 Myr, (a) $>$50-500 Myr, and (c) $>$500-5000 Myr.  
We assume that, in a given volume of space near Earth, stars are
uniformly distributed in age for ages up to $\sim$1 Gyr.  For older
stars one first loses all main-sequence A-type stars - these evolve
off the main-sequence in 1-2 Gyr--followed by loss of F-type
main-sequence stars at ages between $\sim$2 and 4 Gyr \citep{sch92}.

From Figure \ref{age}, there are 26 stars in bin (a), 
74 in bin (b), and 24 in bin (c).  By our assumption of 
equal numbers of stars of any given age in the volume accessible to 
the sensitivity of \iras, the age bin (b) contains 10 times more stars 
in total -- with and without a dusty disk -- than does bin (a).  Since 
bin (b) in Figure \ref{age} contains about 2.8 times the number of 
Vega-like stars as does bin (a), the probability that a star will be 
Vega-like is $\sim$3.5 times greater between ages of 10 and 50 Myr than 
between 50 and 500 Myr.

Similarly, we can estimate the probability that a star in age bin (c)
will be Vega-like.  We ignore for just a moment the loss of A-and F-
type stars in bin (c) as a result of evolution off the main sequence.
In that case, because bin (c) contains 10 times more stars in total -
with and without a dusty disk -- than does bin (b) but fewer Vega-like
stars (24 vs. 74), the probability that a star will be Vega-like in
age bin (b) would be $\sim$30 times greater than in bin (c).  However,
because there is a sequential loss of A- and F-type main-sequence
stars at ages $>$ 1 Gyr, and because these spectral types dominate the
\iras detected 60\,$\micron$ excess stars, we estimate that if a star
has an age appropriate for bin (b), then the probability of its being
Vega-like is only $\sim$10 times (rather than 30 times) greater than
the probability of being Vega-like if its age falls in that of bin 
(b).  Then the probability of any given nearby star in age bin (a)
being Vega-like is $\sim$35 times greater than this probability is in
bin (c).

The preceding discussion pertains to how the probability of being
Vega-like declines with age.  We can estimate the absolute value of
this probability in two ways.  First, two stars in Table~2 are members
of the Hyades (Figure 2: HIP 18975 = VB 160 and HIP 20635 = VB 54) 
although both have cautionary notes and the 60\,$\micron$ excesses 
cannot be regarded as definite until confirmed with additional data.  
\iras could have detected excess 60\,$\micron$ emission comparable to 
$\tau$ = $6 \times 10^{-5}$ at Hyades stars
with V $\lesssim$ 6, which corresponds to a mid-F- type star.
According to Table~1 in \citet{ste95}, 40 Hyades members have a $V$ mag
brighter than 6.  Thus, at an age of 600 Myr, $\leq$5\% of A- through mid-F-
type stars in the Hyades are Vega-like above the 60\,$\micron$ flux
level accessible to \iras.\footnote{\citet{spa01} reported a 60\,$\micron$ 
\iso detection of Hyades member HIP 20261, but at a flux
level, 50 mJy, below the \iras detection limit.}.

Field A-type stars supply a second sample to estimate the probability
that a star will show the Vega-phenomenon.  We find, in essential
agreement with some previous determinations, that \iras detected 60
$\micron$ excess emission at $\sim$20\% of A-type stars with $\tau$
$>$ $10^{-5}$ out to 28 pc (10 of 50 stars) and with $\tau$ $>$ 4 $\times
10^{-5}$ out to 40 pc (22 of 119 stars).  The percentage of F-type
stars that show the Vega phenomenon at comparable levels of $\tau$
appears to be noticeably smaller, but definitive statistics should
wait for results from \textit{Spitzer}.

Notwithstanding the much larger probability of a star being Vega-like 
at young ages, there appears to be very little distinction 
with age in peak $\tau$ seen in Figure \ref{age} and noted in item (1) 
above.  This suggests that the Vega-phenomenon, at least at the 
higher levels of $\tau$ measured by \iras, may be mostly the result 
of occasional large and violent collisional events rather than many 
small-scale, dust-producing events added together.  For example, 
there was a very substantial and recent collisional event at the 
G-type main-sequence star BD +20 307, first detected by \iras at 12 
and 25\,$\micron$ \citep{son05}.

Item (4) noted above might be anticipated in a collisional cascade
model (cf., \citealt{dom03}).  In such a model, collisions grind dust
particles down to smaller and smaller sizes until sufficiently small
particles are blown out of the system by radiation pressure from the
star.  Lower luminosity, later type stars will retain more small
particles in orbit that in total can possess a large emitting area;
thus $\tau$ is increased.  The larger $\tau$ expected for late-type
stars in a \citet{dom03} model is illustrated in their Figure 1\textit{f}.
Earlier, \citet{son01} had suggested that late-type stars display
larger $\tau$ than early-type stars based on the limited data
available to him at that time.

\subsection{Relationship among $\tau$, Disk Mass, Radius, and Stellar Age}

Perhaps the quantities of most interest are disk dust mass, disk radial
extent, and disk evolution with time.  The total mass (\textit{M}) of dust in a
disk may be written:
\begin{equation}
M = \rho N4 \pi a^{3} /3
\end{equation}
where \textit{N} is the total number of grains in the disk and $\rho$ and $a$ are 
the density and radius of a typical grain.  For an optically thin dusty 
ring/shell of characteristic radius \textit{R},
\begin{equation}
\tau = N\pi a^{2}/4\pi R^{2}.
\end{equation}
Then,
\begin{equation}
\tau/M \varpropto 1/\rho aR^{2}.
\end{equation}
Thus, if characteristic grain size and density do not vary much among
various optically thin dust disks, then one expects $\tau$/\textit{M} 
to vary as the inverse square power of the disk radius, \textit{R}.  
Figure \ref{tauMdR} shows this to be approximately the case for dust 
disks with semimajor axes between 10 and 100 AU, where we have taken 
$\tau$ and \textit{R} from Table~2, and disk mass from the submillimeter 
literature.  

The significance of the filled and open symbols in Figure 5 is as 
follows.  The figure was initially prepared containing only the filled 
symbols that represent dust mass determinations based on submillimeter 
data published prior to 2006.  The dashed line was deemed a reasonable 
$R^{-2}$ ``fit'' to these solid symbols and we used it to derive disk dust 
masses for many stars in Table~2 as outlined below.  Then, while the 
present paper was being refereed, a paper presenting submillimeter 
measured masses for six Table~2 stars appeared (HD 14055, 15115, 21997, 
127821, 206893, and 218396; \citealt{wil06}).  These 
six stars appear in our Figure 5 as open symbols, and because they 
lie along the dashed line, they clearly indicate the viability of 
our method.

While recognizing a caveat of statistics of small numbers, relative to
the dashed line the early-type stars preferentially lie somewhat above 
the later type stars .  This difference could be attributed to smaller 
grains around the later-type stars (as discussed in $\S$ 5).  However, 
this model requires that these grains are sufficiently small that they 
are unable to radiate like blackbodies at their temperature and thus, 
at a given distance from the star, are hotter than blackbody grains 
would be at that same distance.

Rather few stars appear in Figure 5 as a direct consequence of the
limited number of published measurements of submillimeter fluxes for
Vega-like stars.  In addition, we plot only stars for which far-IR
excess emission has been measured in at least two wavelengths; for
such stars we can estimate $T_{dust}$ and, thus, $R_{dust}$.

Because $\tau$ is easier to measure (especially with \spit) than is
a submillimeter flux, we use Figure \ref{tauMdR} to derive initial
estimates of dust masses for many stars listed in column (12) of
Table~2.  Combining \iras, \iso and \spit data, all stars 
with masses listed in Table~2 and plotted in Figure 6 have measured 
excess IR emission in at least two wavelengths.  As mentioned in $\S$~4 
above and emphasized in Paper I, the method used to calculate the values 
of R$_{dust}$ listed in Table~2 will sometimes substantially
underestimate the true R$_{dust}$.  Thus, the Table~2 dust 
mass estimates should be regarded with some caution.

The filled symbols in Figure \ref{Md_age} indicate a dust mass measurement 
at submillimeter wavelengths. We expect that stars plotted with ages 
$\lesssim$ 10 Myr still retain significant amounts of orbiting primordial 
dust left over from the star formation process.  Thus, when considering 
the evolution of disk masses in dust, these stars should not be compared 
with the older stars whose dust is of a second generation.  Figure 
\ref{tauMdR} in \citet{naj05}, based solely on submillimeter data, is 
suggestive of dust mass decreasing with time.  However, when stars with ages 
$\lesssim$ 10 Myr are omitted, the remaining submillimeter data are 
consistent with constant average dust mass at stars with ages 
between 30 and 1000 Myr, as suggested by our simple model from Figure 5, 
and the resulting open points are plotted in Figure 6.

\citet{naj05} consider in some detail planet formation models of
Kenyon \& Bromley (2004a,2004b).  According to the discussion in Najita \&
Williams, in these models a wave of planet formation in the disk
propagates outward generating, as time progresses, dusty debris at
successively larger characteristic radii.  According to the models,
for times perhaps as long as 1~Gyr, the total mass in small grains
sensibly remains constant, while, in contrast, the reprocessed
luminosity (i.e., $\tau$) emitted by the collisional debris begins to
decline at a much earlier time ($\lesssim$ 10 Myr).  This is because,
as the wave of planet formation moves outward, grains of a given size
subtend increasingly smaller solid angles the farther they are located
from the star.  Comparing our results (Fig. 4 and 7) with these
models, both a decrease in $\tau$ and an increase in R appears
plausible between 10 and 1000 Myr.  

Figure \ref{tau_AU} is a plot of $\tau$ versus disk radius.  The six stars
with $\tau$ $>$ 10$^{-3}$ all have estimated ages of $\lesssim$ 20
Myr.  Thus, much of their dust may be a remnant of the star formation
process, rather than second generational.  For the other stars, no
correlation is apparent between $\tau$ and \textit{R}.  Although a grain of a
given radius located close to a star will absorb more stellar
radiation than one far away, the lifetime of close-in grains might be
shorter than for distant grains, and these two effects may roughly
cancel, on average.

\subsection{Algol-type binary stars with far-IR excess emission}

An Algol is a binary in which the less massive stellar component
fills its Roche lobe and the other, which does not, is not degenerate 
\citep{bat89}.  Four stars in Table~2 are eclipsing binaries of the 
Algol type, including Algol A itself.  HIP 76267 was long ago 
recognized as a 60\,$\micron$ \iras excess star \citep{aum85}.  The 
\citet{rie05} \spit survey at 24\,$\micron$ included three Algols.  
For HIP 76267, they report a just-significant, 29\%, excess according 
to their criteria (the \spit measured flux must be $>$ 1.25 times 
the expected photosphere to be regarded as significant).  Rieke et 
al.\ also report a 7\% excess at 24\,$\micron$ for Algol A, although 
this does not meet their significance threshold of 25\%.  For HD~40183 
their measured 24\,$\micron$ flux was only 0.88 times the expected 
photosphere.  Although the \iras FSC reports detection of HD 40183 at 
12, 25, and 60\,$\micron$, we see no evidence of an excess at any
wavelength.

The far-IR excess emission at the four Algols might be generated by
free-free and bound-free transitions in ionized gas or by cool dust 
or both.  The Algol-type binary stars are susceptible to emission 
in ionized gas because a small H II region is created around the 
primary star by material transferred from the secondary star.  We 
first consider far-IR emission in an ionized gas disk orbiting a
late B-type primary in Algols listed in Table~3.  We assume an electron
density n$_{e}$ = $10^{10}$ cm$^{-3}$ and disk radius r = $10^{12}$ cm 
\citep{pet89,gui89}.  \citet{cod76} give the flux between 0 and 1100 \AA\ 
received at Earth for the B7 star $\alpha$~Leo.  This translates to $\sim$$2 
\times 10^{44}$ photons s$^{-1}$ emitted by $\alpha$~Leo and capable of ionizing 
hydrogen.  The excitation parameter (\textit{E}), i.e. the number of photons 
per second required to maintain an \ion{H}{2}~region, is
\begin{equation}
E = (4\pi/3) r^{3} n_{e}^{2} \alpha_{B}
\end{equation}
\citep{ost74}.  With $\alpha_{B}$ = 2 $\times$ 10$^{13}$ cm$^{3}$ s$^{-1}$ 
at 10,000 K, and E = 2 $\times 10^{44}$ ionizing photons s$^{-1}$, an
\ion{H}{2}~region with n$_{e}$ = $10^{10}$ cm$^{-3}$ and \textit{r} 
= $10^{12}$ cm can be supported.

Considering the four Algols with SEDs displayed in Figure 9, we assume a
characteristic distance of 30 pc and a characteristic excess flux at 
60\,$\micron$ equal to 0.4 Jy.  The orbiting ionized disk described in the
preceding paragraph would have a 60\,$\micron$ optical depth $\sim$0.2 
\citep{ost74} and could account for this excess flux.  Thus, it is 
plausible that ionized gas, rather than dust, could generate the excess 
far-IR emission in some or even all Algols.

Cool dust might also be present in some of these systems.  The fact that 
Algol itself and HIP~73473 are both triple systems \citep{wor01} may 
supply a clue as to why cool dust is present at all.  In addition to the 
characteristic mass transfer between primary and secondary, analysis 
indicates mass is also lost from Algol systems \citep{bat89}.  If a 
tertiary component is present, then the system could be analogous in 
essential respects to binary post-asymptotic giant branch (post-AGB) 
stars, many of which are known to 
be orbited by a dusty circumbinary disk (e.g., \citealt{wat91}). That is, 
the central object (a single star in the case of the post-AGB stars and 
a binary in the case of Algols), ejects mass, some of which is captured
into a dusty surrounding disk by the gravity of an orbiting companion.

While such a model might apply to Algol A and to HIP~73473,
it need not necessarily apply to other Algols with far-IR excess
emission.  One obvious test would be a search for evidence of a third
star in the HIP~21604 and HIP~76267 systems.

\section{Summary and Conclusions}

The 1983 all-sky \iras far-IR survey yielded a wealth of
information about the properties of cool dust in orbit around main
sequence stars.  However, notwithstanding decades of ground- and
space-based follow-up projects including \iso, as of 2004 when we began
the research reported here, in our opinion, a consistent, convincing 
evolutionary picture of these dusty stars had not been published.  In 
particular, while various researchers had cross-correlated various 
stellar catalogs against the \iras catalog, none had used the 
\hip catalog.  Stellar distances and proper motions provided by the
\hip and Tycho catalogs yield information useful for establishing
ages of dusty stars; reliable ages are essential if correct
evolutionary sequences are to be deduced.  Also, as a consequence of
the rather large \iras beam-size and inadequate attention to
elimination of background confusion, some previous stellar studies
with \iras have suffered from the inclusion of false positive 
far-IR-excess stars.

In the research reported here we have taken special effort to deduce
stellar ages and to eliminate false positives.  Just as it is possible to
deduce many properties of stellar clusters and associations even though
some stars are mistakenly included as members, we trust that our Table~2
\iras sample is clean enough that our conclusions will stand the test of
time.  Nonetheless, because ages of nearby field stars are notoriously
difficult to estimate accurately and because of limitations with the \iras
database, we recognize that some entries in the tables and figures
presented in this paper will be in error.

\iras was most effective for the study of luminous B- and A-type main-sequence
stars.  In agreement with some earlier studies, we find that \iras detected
excess emission at 60\,$\micron$ from about 20\% of nearby A-type stars.  
This percentage will certainly rise as the A stars are examined with
far-IR photometers more sensitive than those aboard \iras.  In particular, 
we find that about 10\% of stars of various spectral classes are revealed 
to display far-IR dust emission at brightness levels between \iras and 
\spit sensitivities.  Although this 10\% subsumes stellar age, spectral 
types, and distance from Earth and thus is potentially subject to selection 
effects, it is consistent with the well-defined TW Hydra association 
sample of \citet{low05}.  Using heterogeneous samples, \citet{smi06} 
and \citet{bry06} also reported about 10\% of stars show dust excess in 
the MIPS 70\,$\micron$ band, but below \iras sensitivity.

From their analysis of \iso data sets, especially the volume-limited
sample of \citet{hab01}, \citet{dec03} deduced that the percentage of 
stars with detectable 60\,$\micron$ emission diminishes with age.  
However, the small data set of Habing et al. (2001) and difficulties 
with estimating stellar age, precluded a meaningful quantitative result 
in our opinion.  With our larger and more robust database we can derive 
that the probability of 60\,$\micron$ excess emission detectable with 
the sensitivity of \iras is about 35 times larger for 
stars with ages in the range 10-50 Myr compared to such stars with ages
$>$ 500 Myr within the volume within 120 pc of Earth.

While it is generally agreed that measurements at submillimeter
wavelengths are best for the derivation of dust masses, by means of a
simple model that relates submillimeter and far-IR fluxes, we are able
to derive dust masses for numerous stars that lack submillimeter
data. These masses lie in the range between 0.0005 and 0.5 \textit{M}$_{\earth}$.
For stars with ages between 30 and 1000 Myr, these dust masses appear
to depend little, if at all, on age.  Based on Figure 5, and as 
described in $\S$~5.1, our model indicates that far-IR data can be used, 
quite reliably, to predict a submillimeter flux and, thus, a disk 
dust mass.  As a consequence, disk dust masses can generally be 
derived based solely on \spit data provided that excess flux is 
measured at two or more well-separated wavelengths with MIPS and/or 
the Infrared Spectrograph (IRS).

Four Algol binary stars appear to display excess emission at 60\,$\micron$
wavelength, although the existence of the excess is perhaps not compelling
in all cases.  We considered models in which the emission is generated by
free-free and bound-free emission in orbiting ionized gas or by orbiting
dust particles, dust perhaps associated with a tertiary (third)  stellar
component.  Future studies will be required to clarify the dominant
physical mechanism(s) involved.

Additional results of our study include:  (1) Peak $\tau$ ($\sim10^{-3}$) 
does not vary much at all ages later than $\sim$10\,Myr; this might 
be because occasional catastrophic dust-generating 
events can occur at any age. (2) The spread of measured $\tau$ at 
ages $\sim$10 Myr is about a factor of 10, increasing to about 100 
at later ages; given the measured peak $\tau$ (item 1) and \iras 
threshold ($\sim10^{-5}$), the measured spread of $\tau$ cannot be greater 
than 100. (3) At any given age late-type stars tend to have the largest 
$\tau$.  (4) For stars with ages between 10 and 1000 Myr, the mean $\tau$ 
of stars with \iras detectable far-IR excess emission declines in proportion 
to (age)$^{0.7}$.   (5) For early-type stars between ages of $\sim$10 and 
100 Myr, the typical radius of a dusty debris disk appears to be smaller 
than for stars with ages between 100 Myr and 1 Gyr. (6) The very largest 
taus ($>$ $10^{-3}$) are associated only with disks that have relatively 
small radii.  (7) \iras detected excess 60\,$\micron$ emission from 
$\sim$20\,\% of nearby A-type stars.  (8) Four Algol-type eclipsing 
binaries, including Algol A itself, display 60\,$\micron$ emission, 
generated by free-free and bound-free transitions in ionized gas, by 
dust grains, or by both.  (9) Gl~803 (AU~Mic, 12 Myr old) is the only 
M-type, non-T Tauri, \hip dwarf star to display 60\,$\micron$ 
excess emission in the \iras Catalogs.
\acknowledgments

We thank the referee for his/her constructive comments that helped to improve 
this paper.  We also thank Travis Barman for providing a customized set 
of NextGen Pheonix models of stellar atmospheres, and M. Jura for helpful 
comments.  This research has made use of the VizieR catalogue access tool, CDS, 
Strasbourg, France and of data products from the Two Micron All Sky Survey 
(The latter is a joint project of the University of Massachusetts and the 
Infrared Processing and Analysis Center/California Institute of Technology, 
funded by the National Aeronautics and Space Administration and the 
National Science Foundation).  We acknowledge a NASA grant NAG-13067 
for financial support.  

\clearpage
\appendix
\section{Appendix material}
\begin{deluxetable}{cccccc}
\tablenum{A4}
\tabletypesize{\scriptsize}
\tablecaption{List of Rejected Sources}
\tablewidth{0pt}
\tablehead{
\colhead{}              &\colhead{}           &\colhead{\iras}       &
\colhead{Contamination} &\colhead{Additional} &\colhead{Reason of}  \\
\colhead{HIP}           &\colhead{HD}         &\colhead{Source}     &
\colhead{Source} &\colhead{Data Source} &\colhead{Rejection\tablenotemark{\dag}}  \\
}
\startdata
   1468 &   1407 & F00157+1907 & UGC 00169 & NED                      & 1   \\
   2021 &   2151 & F00235-7731 &  &                                   & 2   \\
   8102\tablenotemark{\ddag} &  10700 & F01416-1611 &  & MIPS	                      & 3   \\
   8796 &  11443 & F01502+2919 &  &                                   & 2   \\
   8817 &    ... & F01506+2312 & 2MASX J01532347+2327067 & NED        & 1   \\
   9236 &  12311 & F01572-6148 &  &                                   & 2   \\
  12843 &  17206 & F02427-1846 &  & MIPS                              & 3   \\
  13847 &  18622 & F02563-4030 &  &                                   & 2   \\
  14897 &  20010 & F03095+1351 &  &                                   & 2   \\
  15197 &  20320 & F03134-0900 &  &                                   & 2   \\
  16276 &  20110 & F03190+8352 & HIP 16267 &                          & 4   \\
  17378 &  23249 & F03408-0955 &  &                                   & 2   \\
  17439 &  23484 & F03423-3826 &  & \iso                               & 5   \\
  17531 &  23338 &  03421+2418 &  &                                   & 5,6 \\
  17573 &  23408 & F03428+2412 & NGC 1432 & NED                       & 1   \\
  17579 &  23432 &  03429+2423 &  &                                   & 5,6 \\
  17608 &  23480 & F03433+2347 &  &                                   & 5,6 \\
  17921 &  23950 & F03469+2205 &  &                                   & 5,6 \\
  21010 &  28447 & F04273+2800 & 2MASX J04302705+2807071 & NED        & 7   \\
  22449 &  30652 & F04471+0652 &  &                                   & 2   \\
  23818 &  33095 & F05049-1927 &  &                                   & 1   \\
  25110 &  33564 & F05142+7911 & \iras F05142+7911 & MIPS              & 1   \\
  25732 &  36150 &  05271-0050 &  &                                   & 5   \\
  27100 &  39014 & F05446-6545 &  &                                   & 5   \\
  28360 &  40183 & F05558+4456 &  &                                   & 2   \\
  30252 &  44958 & F06207-5112 &  &                                   & 8   \\
  32277 &    ... & F06407+4040 & HIP 32275 &                          & 4   \\
  32349 &  48915 &  06429-1639 &  &                                   & 2   \\
  32435 &  53842 & F06539-8355 &  & MIPS                              & 9   \\
  34473 &  55864 & F07091-7024 &  &                                   & 2   \\
  35457 &  56099 & F07149+5913 &  & MIPS                              & 9   \\
  35789 &  58853 & F07225-6432 &  & \iras F07225-6432                  & 1   \\
  37279 &  61421 & F07366+0520 &  & MIPS                              & 2   \\
  40167 &  68255 & F08093+1747 &  &                                   & 10  \\
  42913 &  74956 &  09433-5431 &  & MIPS                              & 3   \\
  43100 &  74738 & F08436+2856 & HIP 43103 &                          & 4   \\
  44923 &  78702 & F09067-1807 &  & MIPS                              & 5   \\
  44915 &  78752 & F09068-2844 &  &                                   & 5   \\
  45238 &  80007 & F09126-6930 &  &                                   & 2   \\
  46853 &  82328 & F09294+5154 &  &                                   & 2   \\
  46984 &  82821 & F09319+0346 & 2MASX J09343627+0332421 & NED        & 1   \\
  49641 &  87887 & F10053-0007 &  & MIPS                              & 3   \\
  49669 &  87901 & F10057+1212 &  &                                   & 2   \\
  54835 &  97455 & F11107+5541 & SBS 1110+556 & NED                   & 1   \\
  57583 &    ... & F11457-2150 &  &                                   & 11   \\
  57757 & 102870 & F11481+0202 &  &                                   & 2   \\
  57759 & 102902 & F11482-3252 & Unknown galaxy & NED                 & 12   \\
  58001 & 103287 & F15512+5358 &  &                                   & 2   \\
  58364 & 103913 &  11554+2524 &  & NED                               & 1   \\
  59307 & 105686 & F12074-3425 & GdF J1209598-344142 & NED            & 1   \\
  60112 & 107228 & F12171+0549 & NGC 4266 & NED                       & 1   \\
  60902 & 108653 & F12263+0126 & SDSS J122856.95+010907.4 & NED       & 1   \\
  61932 & 110304 & F12387-4841 &  &                                   & 2   \\
  61941 & 110379 & F12390-0110 &  &                                   & 2   \\
  61947 &    ... & F12394+4319 & 2MASX J12414864+4302494 & NED        & 1   \\
  62956 & 112185 & F12518+5613 &  &                                   & 2   \\
  63973 & 113767 &  13036-4924 & NGC 4945A & NED                      & 1   \\
  65109 & 115892 & F13177-3627 &  & MIPS                              & 2   \\
  65378 & 116656 & F13219+5511 &  &                                   & 2   \\
  66249 & 118098 & F13321-0020 &  & \iso                               & 2   \\
  67927 & 121370 & F13522+1838 &  &                                   & 2   \\
  68160 & 121812 & F13549+2336 &  &                                   & 6   \\ 
  68380 & 122106 & F13571-0318 & APMUKS(BJ) B135713.55-031828.8 & NED & 1   \\
  70497 & 126660 & F14235+5204 &  &                                   & 2   \\
  72339 & 130322 & F14449-0004 & APMUKS(BJ) B144458.55-000415.4 & NED & 1   \\
  72659 & 131156 & F14491+1918 &  &                                   & 2   \\
  75039 & 136580 & F15182+4109 & 2MASX J15200834+4059114 & NED        & 1   \\
  75118 & 136407 & F15182-1522 &  & MIPS	                      & 3   \\
  76641 & 139907 & F15374+4401 & UGC 09959 & NED                      & 1   \\
  77634 & 141556 &  15477-3328 &  &                                   & 6   \\
  78072 & 142860 & F15541+1548 &  & MIPS                              & 2   \\
  78527 & 144284 & F16009+5841 &  &                                   & 2   \\
  78594 & 143840 & F16001-0440 &  & MIPS                              & 10   \\
  79807 & 147094 & F16159+5229 & 2MASX J16171300+5222153 & NED        & 1   \\
  81693 & 150680 & F16393+3141 &  &                                   & 2   \\
  83137 & 153377 & F16567-0136 &  & MIPS                              & 3   \\
  83343 &    ... & F16599+2300 &  &                                   & 13  \\
  84696 & 156635 & F17162-0245 &  &                                   & 1   \\
  85104 &    ... & F17223+4811 &  &                                   & 9   \\
  85576 & 158373 & F17265-0957 &  & \iso                               & 6   \\
  85790 & 159139 &  17299+2826 & CGCG 170-036 & NED                   & 1   \\
  86032 & 159561 & F17326+1235 &  &                                   & 2   \\
  86974 & 161797 & F17444+2744 &  &                                   & 2   \\
  87815 & 164330 & F17559+6236 &  & \iso                               & 6   \\
  89937 & 170153 & F18220+7242 &  &                                   & 2   \\
  92683 & 174966 &  18505+0141 &  &                                   & 14  \\
  93371 & 176270 & F18576-3708 & IC 4812 & NED                        & 1   \\
  93449 &    ... & F18585-3701 & NGC 6729 & NED                       & 1   \\
  98025 & 189207 & F19544+6227 &  & MIPS                              & 3   \\
  98433 & 189478 &  19575+0647 &  &                                   & 6   \\
  99240 & 190248 & F20039-6619 &  &                                   & 2   \\
 104206 & 199391 & F20593-8053 &  & MIPS                              & 3   \\
 105090 & 202560 & F21141-3904 &  & MIPS                              & 2   \\
 105858 & 203608 & F21223-6535 &  & MIPS                              & 2   \\
 106368 & 204942 & F21297-2422 & APMUKS(BJ) B212943.47-242303.3 & NED & 1   \\
 107556 & 207098 & F21442-1621 &  &                                   & 2   \\
 108594 &    ... & F21563-6220 & APMUKS(BJ) B215622.59-622020.9 & NED & 1   \\
 108870 & 209100 & F21598-5700 &  & MIPS                              & 2   \\
 111544 & 214168 & F22335+3921 & HIP 111546 &                         & 5   \\
 111558 &    ... & F22330-5154 & ESO 238-IG 019 & NED                 & 1   \\
 114996 & 219571 & F23145-5830 &  & \iso                               & 2   \\
 118182 &    ... & F23558+5106 & HIP 118188 &                         & 5   \\
 118268 & 224617 & F23567+0634 &  &                                   & 6   \\
\enddata
\tablenotetext{\dag}{
1. There exists a nearby extended source within 3\,$\sigma$ \iras positional error ellipse. \\
2. SED shows that \iras 60\,$\micron$ or MIPS 70\,$\micron$ detection falls on the 
   stellar photosphere. \\
3. No source was detected at the expected stellar position in MIPS 70\,$\micron$ image. \\
4. There exists a 2nd bright star within 3\,$\sigma$ \iras positional error ellipse. \\
5. \iras 60\,$\micron$ excess is likely caused by cirrus contamination. \\
6. This star, a member of the Pleiades cluster, is likely contaminated by cirrus \citep{kal02}. \\
7. \iras SCANPI shows 1$'$ offset in inscan direction where the listed galaxy is located. \\
8. 3\,$\sigma$ \iras positional error ellipse does not include the target star. \\
9. \citet{moo06} rejected this star based on their \spit MIPS observation. \\
10. Infrared excess had $<$ 2.5\,$\sigma$ detection at \iras 60\,$\micron$ band (see $\S$~2 for the definition of $\sigma$).\\
11. \iras FSC long format indicates a large offset between 60\,$\micron$ and 12\,$\micron$ positions. \\
12. \iras SCANPI shows 30\,$\arcsec$ offset in inscan direction where the listed galaxy is located. \\
13. \spit MIPS 70\,$\micron$ image shows extended emission. \\
14. There exists a huge background galaxy behind this star. \\
}
\tablenotetext{\ddag}{Both \iras \& \iso reported excess emission at 60\,$\micron$, and \citet{gre04} 
reported excess emission at 850\,$\micron$.  However, \spit MIPS observation shows stellar 
photosphere detection at 70\,$\micron$.}
\end{deluxetable}


\clearpage

\clearpage
\begin{figure}
\includegraphics[scale=1.0]{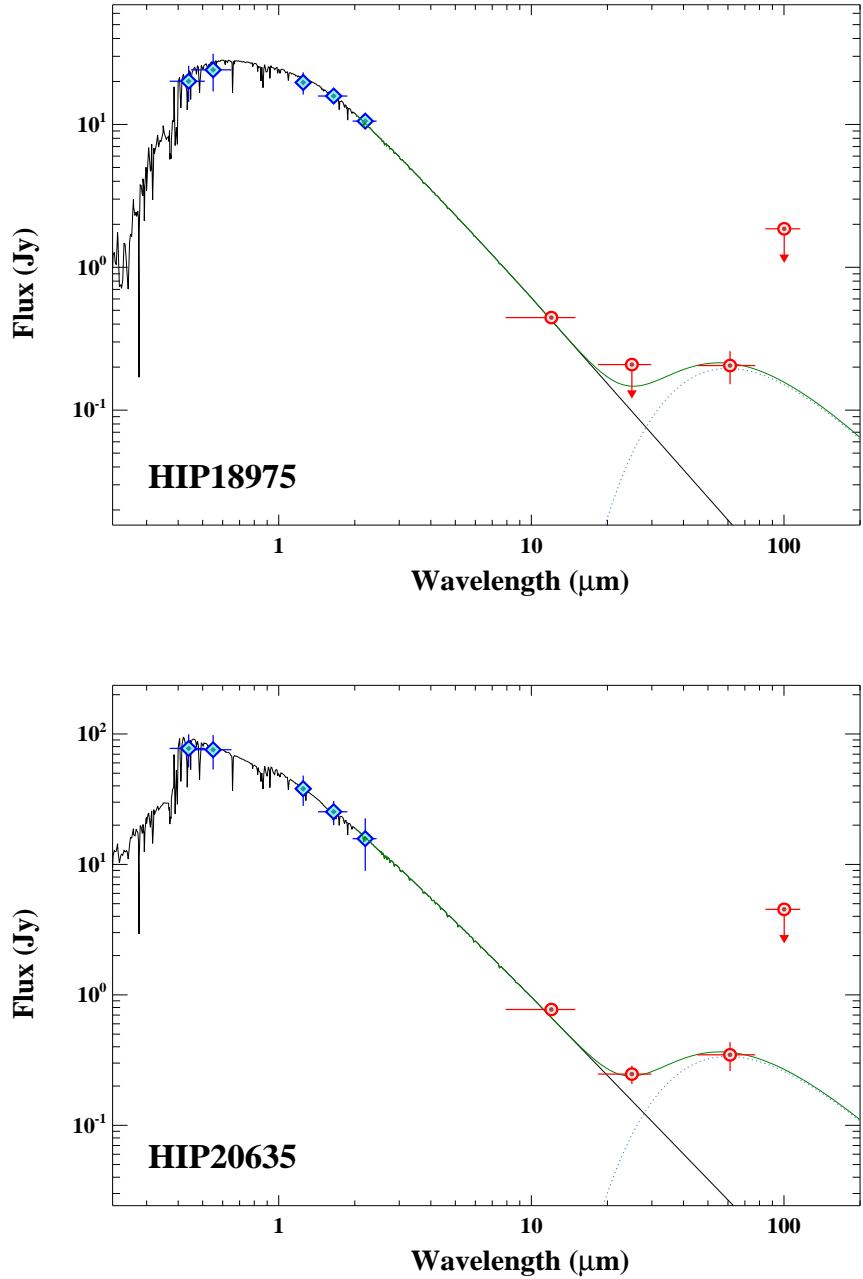}
\caption{Spectral Energy Distribution (SED) of Hyades stars.  Fitting 
parameters (e.g., R$_{\star}$, T$_{\star}$, R$_{dust}$, T$_{dust}$) of each 
star are given in Table~2 which also gives cautionary notes so that the 
apparent 60\,$\micron$ excesses seen in these SEDs cannot be regarded 
as definite until confirmed with additional data.  SEDs of the remaining 
IR-excess stars are available in the electronic edition.}
\label{hyadesSED}
\end{figure}

\clearpage
\begin{figure}
\plotone{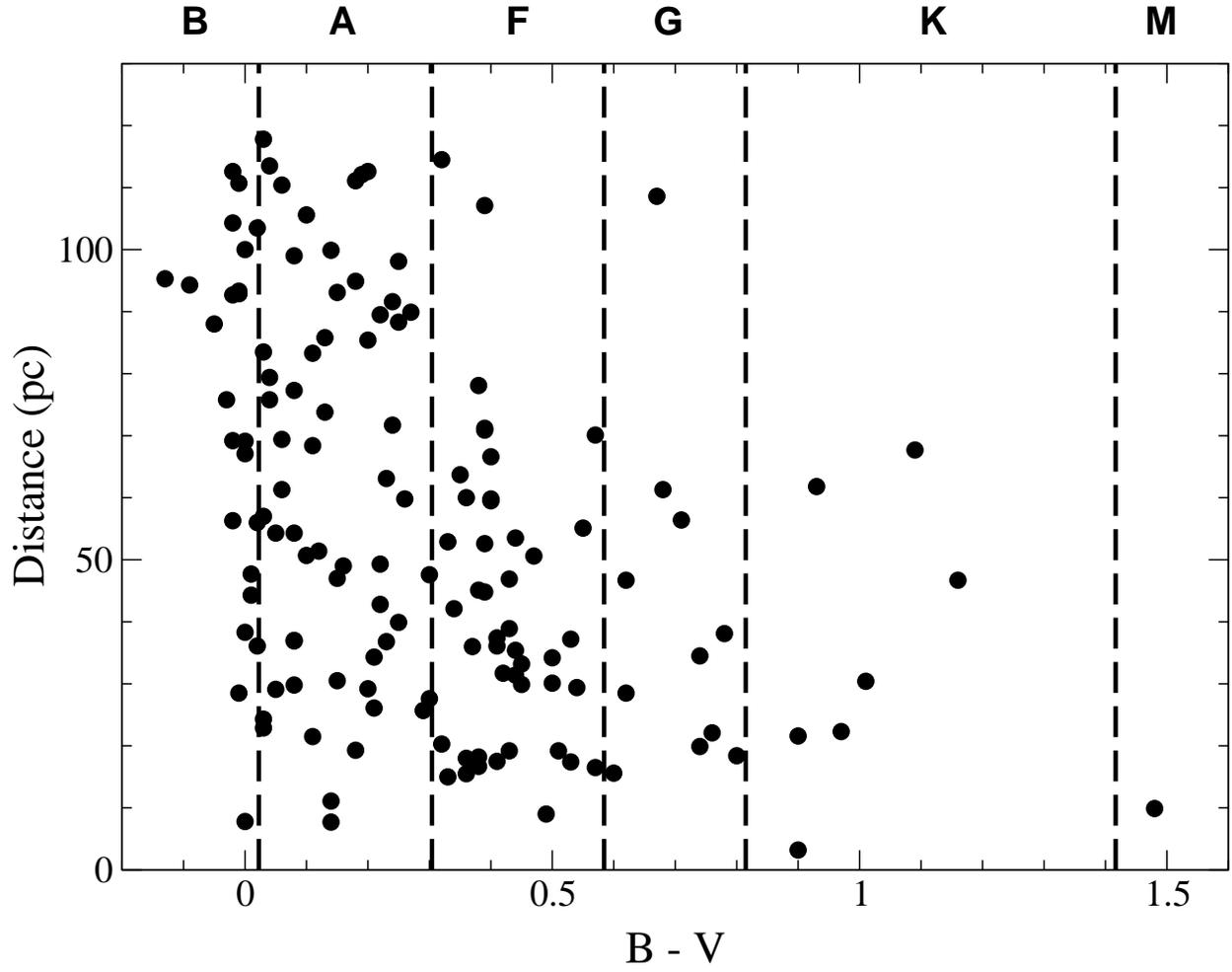}
\caption{Distribution of our 146 candidate excess stars in distance from 
Earth as a function of \bv.  As reported previously, early type stars 
dominate the \iras debris disk systems.}
\label{dist_bv}
\end{figure}

\clearpage
\begin{figure}
\includegraphics[scale=0.54,angle=90]{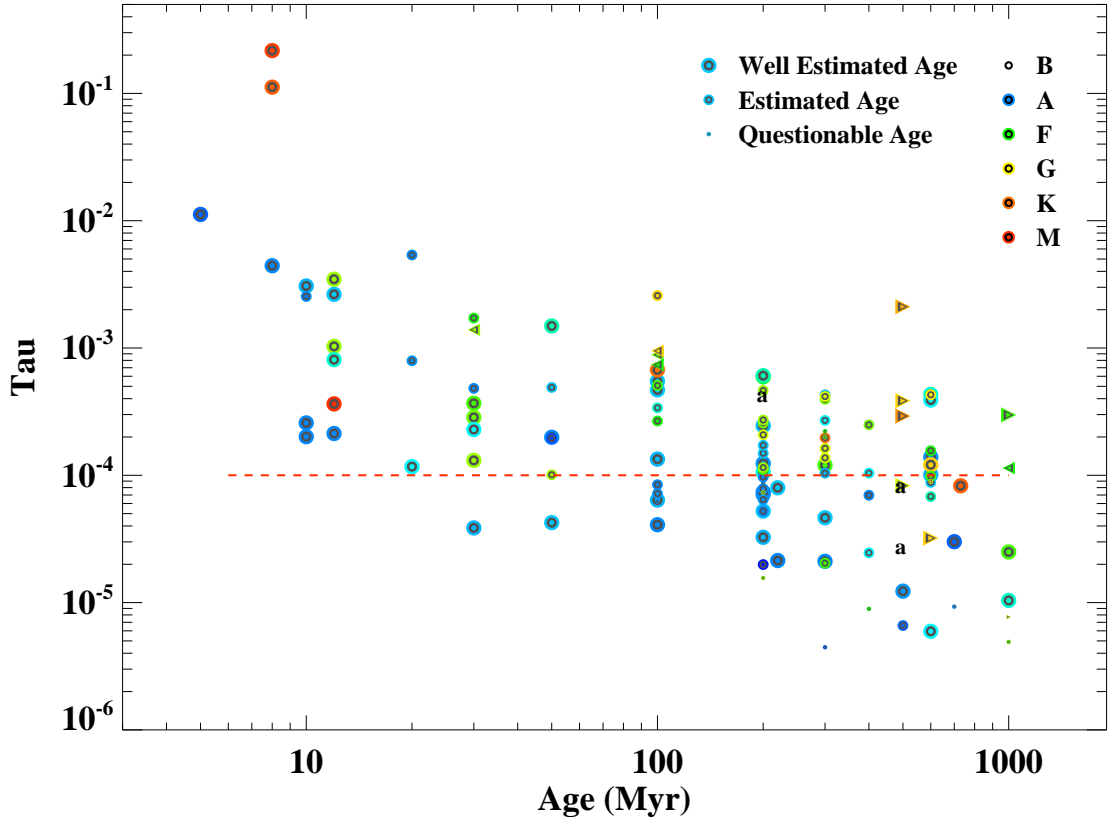}
\caption{$\tau$ as a function of stellar age.  Plotted lower case 
``a'' are Algol-type stars.  Well estimated age, 
estimated age, and questionable age correspond respectively to zero, 
one, and two question marks in column (13) of Table~2.  Stars with 
cautions noted in Table~2 for possible contamination are not plotted 
in the figure.}
\label{age}
\end{figure}

\clearpage
\begin{figure}
\includegraphics[scale=0.86]{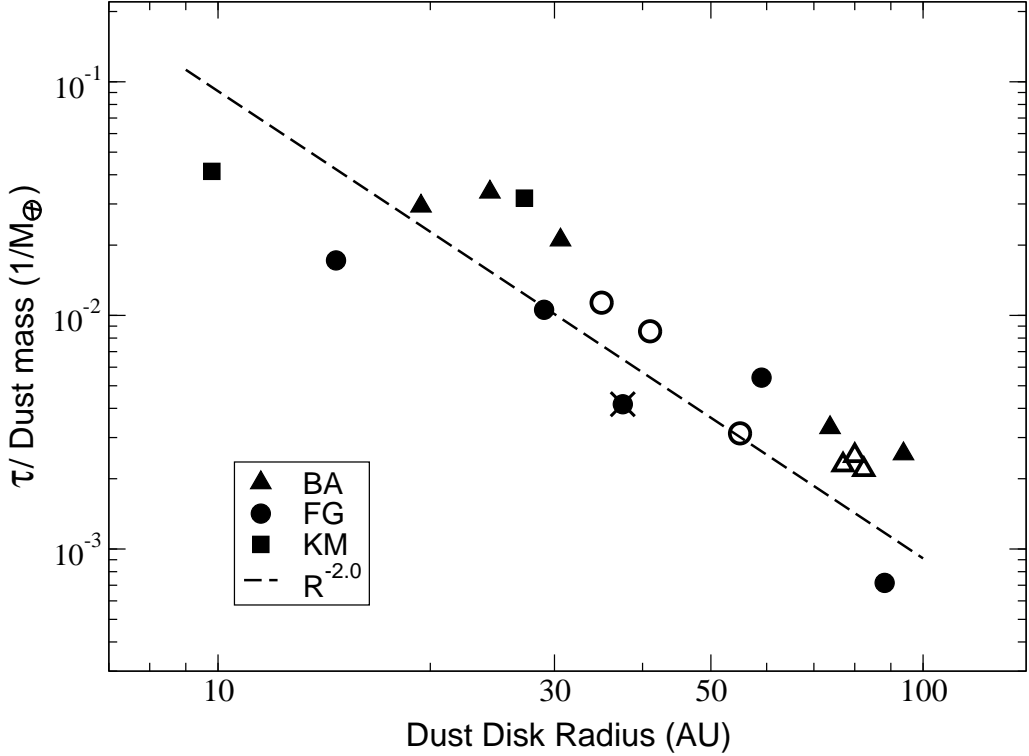}
\caption{$\tau/M_{dust}$ as a function of dust disk radius (AU).  $M_{dust}$, 
given in Earth masses ($M_{\earth})$, is derived from submillimeter 
measurements reported in the published literature.  The filled and open 
symbols represent dust mass determination based on submillimeter data 
published prior to 2006 and during 2006, respectively.  The dashed line 
has slope, $R^{-2}$, but is not a formal ``best fit'' to the data points.  
See $\S$~5.1 for further 
discussion.  To achieve consistency among data reported in various 
published papers, all masses given in the plot have been normalized 
(by us) to have a dust opacity of 1.7 $cm^{2} g^{-1}$ at 850\,$\micron$ 
and dust temperature as given in our Table 2.  However, uncertainties 
in the 850\,$\micron$ dust opacity caused by different grain sizes and 
compositions can result in the over- or underestimate of dust mass by a 
factor of three or so (e.g., \citealt{pol94,bec00}).  Meanwhile, the 
relative masses of the various submillimeter determinations might be 
better constrained than their absolute values if each star has reasonably 
similar dust.  In the Figure, the relative masses are probably 
trustworthy to about a factor of two.  All stars plotted have measured
far-IR excess emission in at least two wavelengths.  $\tau$ for one 
star (HD 104860) is from \iso, not \iras, and is marked ``$\times$''. }
\label{tauMdR}
\end{figure}
\clearpage

\clearpage
\begin{figure}
\includegraphics[scale=0.54,angle=90]{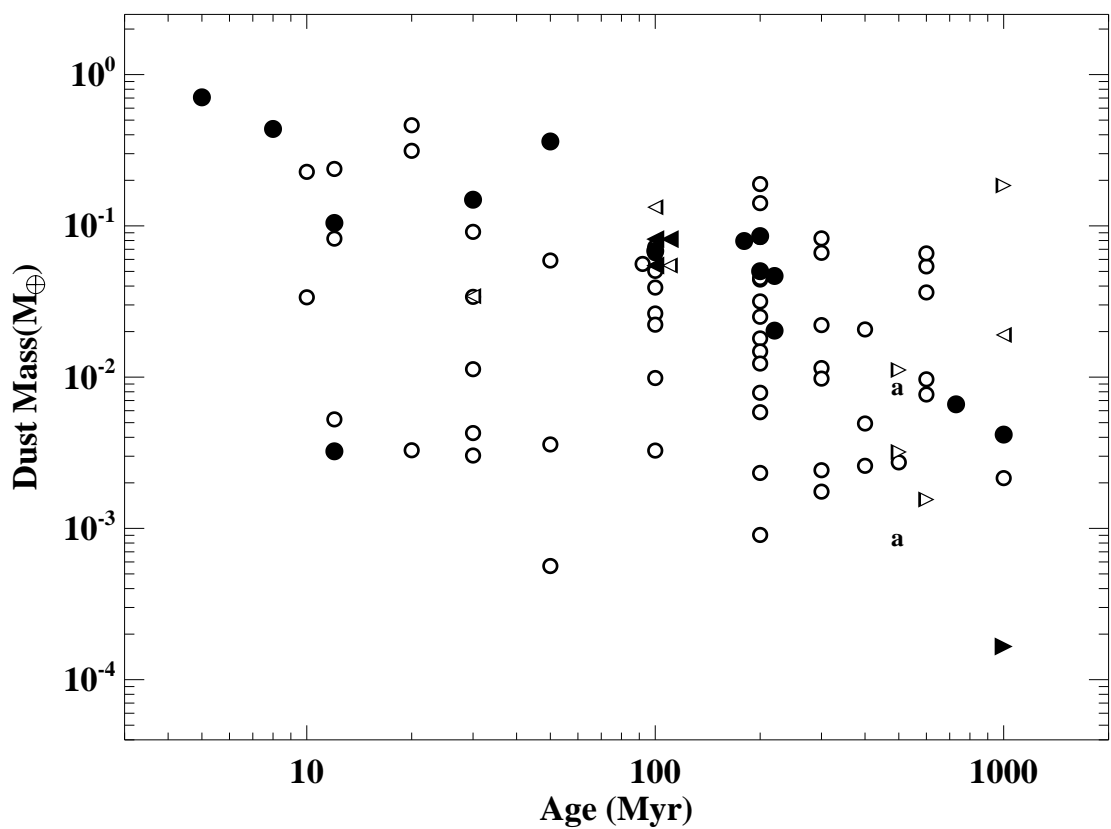}
\caption{M$_{dust}$ as a function of stellar age.  Solid symbol
depicts M$_{dust}$ obtained from submillimeter measurements while open
symbol represents M$_{dust}$ derived from Figure \ref{tauMdR} (see
$\S$~5.1).  All stars plotted have measured far-IR excess emission in at
least two wavelengths and R$_{dust}$ between 9 and 100 AU.  Stars with 
cautions noted in Table~2 for possible contamination are not plotted 
in the figure.  Two Algol-type stars are plotted with lower case ``a'' 
(although their IR excess may not be due to dust particles, see $\S$~5.2).}
\label{Md_age}
\end{figure}

\clearpage
\begin{figure}
\includegraphics[scale=0.54,angle=90]{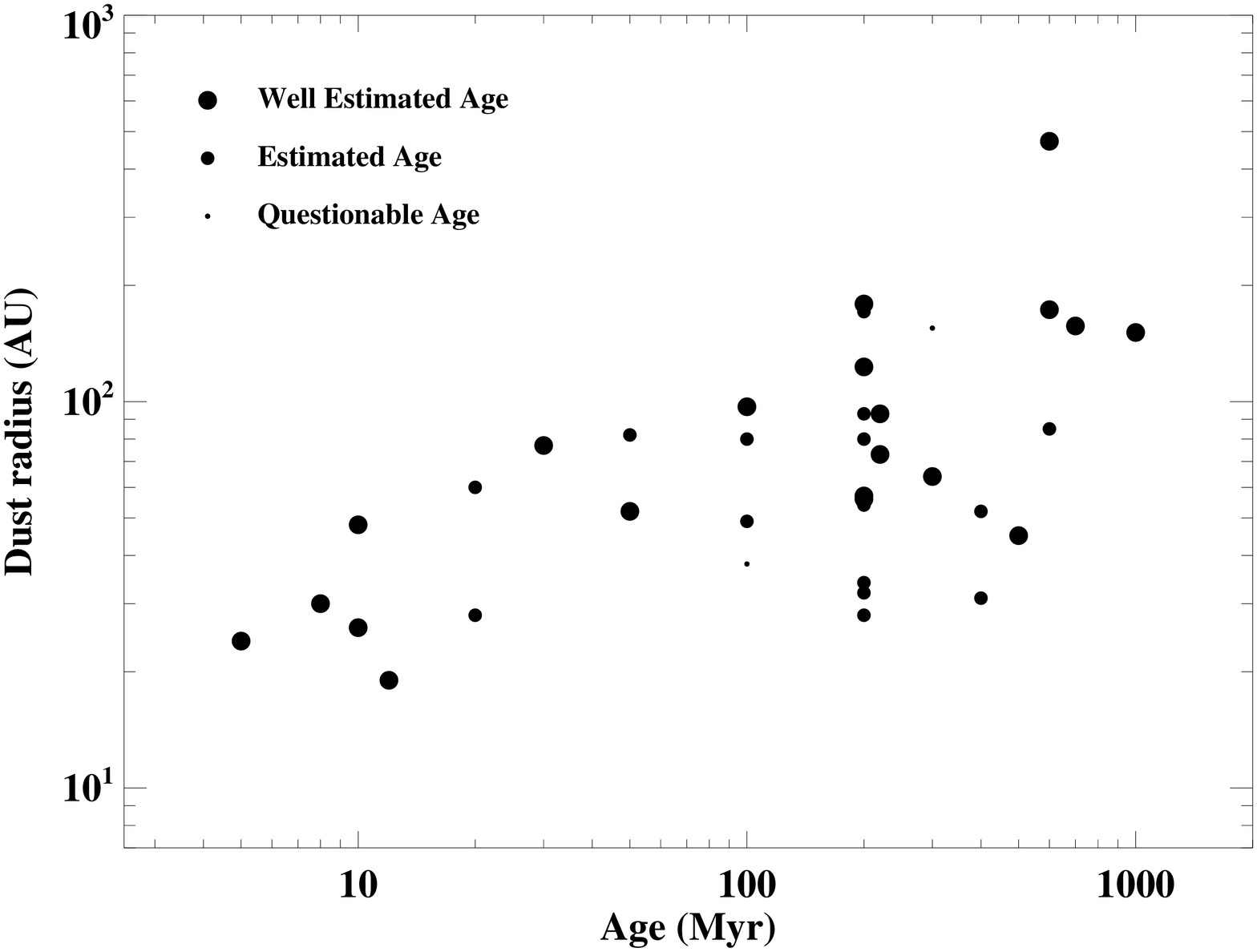}
\caption{Dust radii of early type IR-excess stars (B \& A) as a function 
of stellar age. All stars plotted have measured far-IR excess emission
in at least two wavelengths.  Stars with 
cautions noted in Table~2 for possible contamination are not plotted 
in the figure.}
\label{dR_age}
\end{figure}

\clearpage
\begin{figure}
\includegraphics[scale=0.54,angle=90]{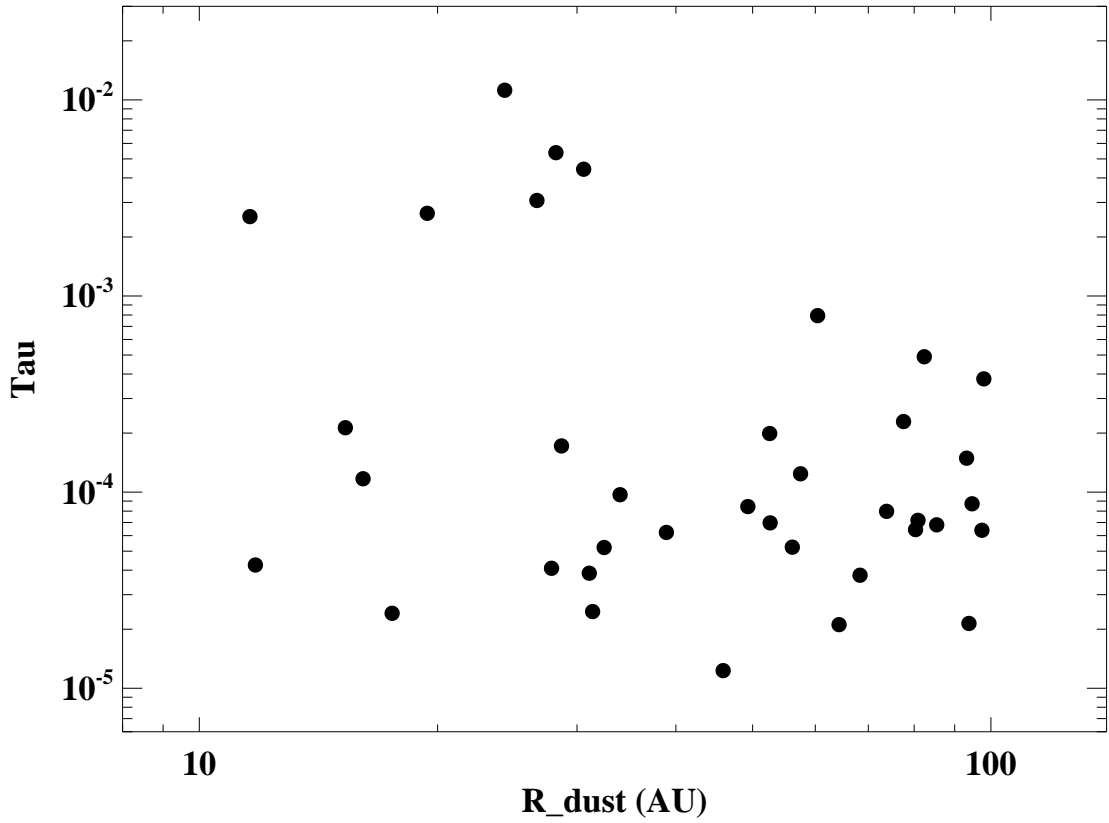}
\caption{$\tau$ of early type IR-excess stars (B \& A) as a function of 
dust radii. All stars plotted have measured far-IR excess emission in at 
least two wavelengths.  Stars with 
cautions noted in Table~2 for possible contamination are not plotted 
in the figure.}
\label{tau_AU}
\end{figure}

\clearpage
\begin{figure}
\epsscale{1.0}
\plotone{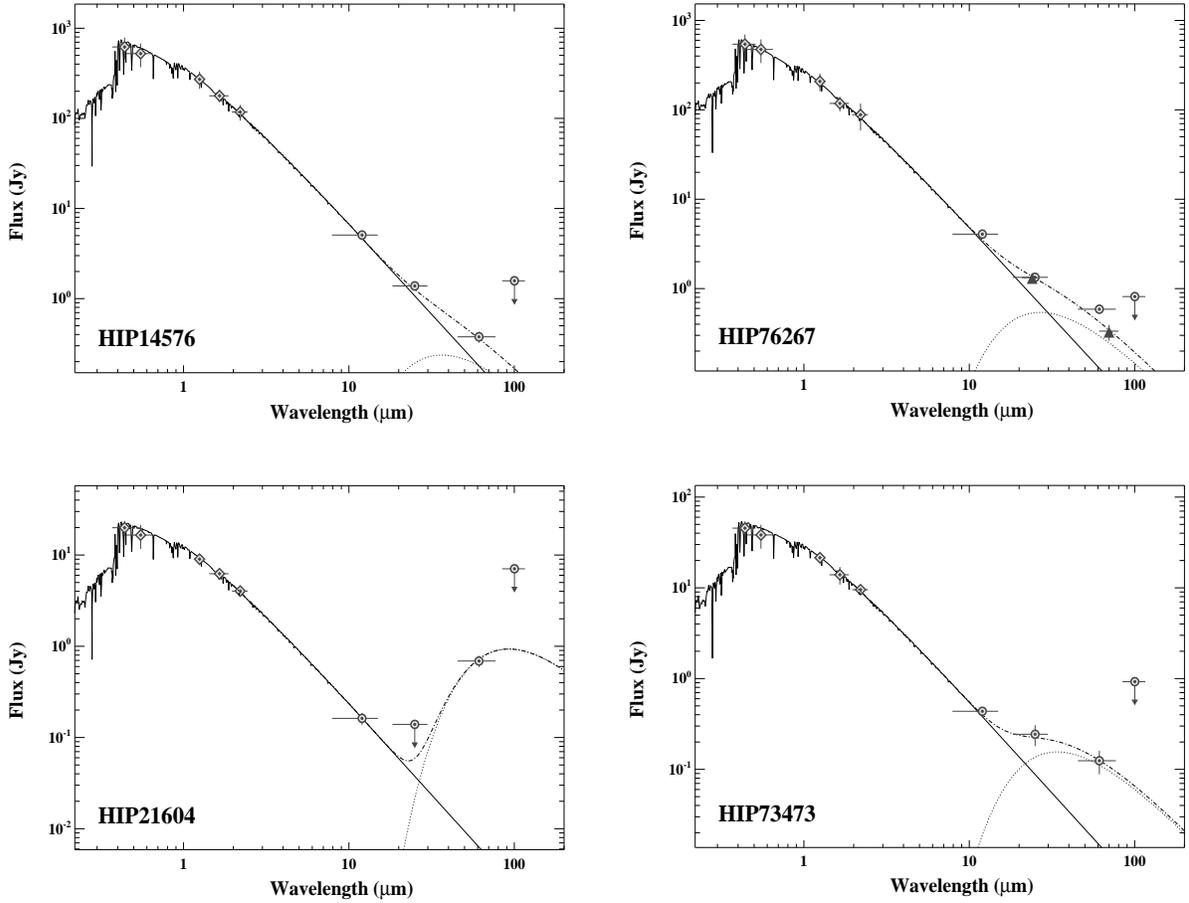}
\caption{Spectral Energy Distribution (SED) of Algol-type stars.  For
HIP 76267 the filled triangle data points at 24 and 70\,$\micron$ are
from \citet{rie05}.  Fitting parameters (e.g., R$_{\star}$, T$_{\star}$,
R$_{dust}$, T$_{dust}$) of each star are given in Table~2.  However,
the far-IR emission might be generated by ionized gas (see $\S$~5.2).}
\label{algolSED}
\end{figure}

\clearpage
\pagestyle{plain}
\begin{deluxetable}{cccccc}
\tablenum{1}
\tablecolumns{6}
\tablecaption{\hip Class I \& II Pre-main Sequence Stars within 120 pc}
\tablewidth{0pc}
\tablehead{
\colhead{HIP}                &\colhead{HD}               &\colhead{Other}      &
\colhead{Sp. Type}           &\colhead{V}             &\colhead{Distance}   \\
\colhead{}                   &\colhead{}                 &\colhead{}           &
\colhead{}               &\colhead{(mag)}                 &\colhead{(pc)}                 
}
\startdata
17890 & 275877 & XY Per   & A2IIev    & 9.44  & 120.0 \\
23873 &	240764 & RW Aur A & G5V:e...  & 10.3   & 70.5 \\
26295 &	36910  & CQ Tau   & F2IVe     & 10.7   & 99.5 \\
56354 &	100453 & ...      & A9Ve      & 7.79   & 111.5 \\
56379 &	100546 & KR Mus   & B9Vne     & 6.70   & 103.4 \\
58520 &	104237 & DX Cha   & A:pe      & 6.60   & 116.1 \\
82323 &	...    & V1121 Oph  & K5      & 11.25  & 95.1 \\
\enddata
\end{deluxetable}

\clearpage
\pagestyle{empty}
\begin{deluxetable}{c@{}c@{}c@{}c@{}c@{}c@{}c@{}c@{}c@{}c@{}c@{}c@{}c@{}c@{}c@{}c}
\tablenum{2}
\tabletypesize{\scriptsize}
\rotate
\tablecaption{Stars with Dusty Debris Disks \label{tbl-1}}
\tablewidth{0pt}
\tablehead{
\colhead{}           &\colhead{}           &\colhead{}           &
\colhead{V}          &\colhead{D}          &\colhead{$R_{\star}$} &
\colhead{$T_{\star}$} &\colhead{$T_{dust}$} &\colhead{$R_{dust}$} &
\colhead{angle}      &\colhead{}           &\colhead{Dust mass}  &
\colhead{age}        &\colhead{Age}        &\colhead{Dust Excess}&
\colhead{} \\
\colhead{HIP}        &\colhead{HD}         &\colhead{Sp. Type}   &
\colhead{(mag)}      &\colhead{(pc)}       &\colhead{($R_\odot$)}&
\colhead{(K)}        &\colhead{(K)}        &\colhead{(AU)}       &
\colhead{(arcsec)}   &\colhead{$\tau$}     &\colhead{$M_{\oplus}$} &
\colhead{(Myr)}      &\colhead{Method\tablenotemark{\dag}}          &
\colhead{Confirmation}&\colhead{Notes\tablenotemark{\alpha}} \\
\colhead{(1)}        &\colhead{(2)}        &\colhead{(3)}        &
\colhead{(4)}        &\colhead{(5)}        &\colhead{(6)}        &
\colhead{(7)}        &\colhead{(8)}        &\colhead{(9)}        &
\colhead{(10)}       &\colhead{(11)}       &\colhead{(12)}       &
\colhead{(13)}       &\colhead{(14)}       &\colhead{(15)}       &
\colhead{(16)} \\
}
\startdata
    746 &    432 & F2III-IV &   2.3 &  16.7 & 3.36 &  7200 &  120 &   28 &  1.68 & 2.50E-05 & 2.15E-03 &    1000 &      a,b,c & ...	 &\\
   1185 &   1051 &    A7III &   6.8 &  88.3 & 1.87 &  8000 &   40 &  173 &  1.97 & 4.32E-04 &          &     600 &        a,d & ...	 &\\
   4267 &   5267 &     A1Vn &   5.8 & 112.6 & 2.67 & 10000 &   85 &   86 &  0.76 & 8.77E-05 &          &     200 &        a,d & ...	 &1,2,3\\
   5626 &   6798 &      A3V &   5.6 &  83.5 & 2.25 & 10000 &   75 &   93 &  1.12 & 1.49E-04 & 1.41E-01 &    200? &        a,d & ...	 &\\
   6686 &   8538 &     A5Vv &   2.7 &  30.5 & 3.90 &  8400 &   85 &   88 &  2.90 & 5.95E-06 &          &     600 &        a,d & ...	 &\\
   6878 &   8907 &       F8 &   6.7 &  34.2 & 1.19 &  6600 &   45 &   59 &  1.74 & 2.08E-04 & 3.84E-02\tablenotemark{\ddag} &    200? &      a,b,c & MIPS/\iso &\\
   7345 &   9672 &      A1V &   5.6 &  61.3 & 1.66 & 10000 &   80 &   60 &  0.99 & 7.94E-04 & 3.13E-01 &     20? &        ZFK & ...	 &4\\
   7805 &  10472 &   F2IV/V &   7.6 &  66.6 & 1.28 &  7000 &   70 &   30 &  0.45 & 3.68E-04 & 3.39E-02 &      30 &        ZSW & MIPS	 &\\
   7978 &  10647 &      F8V &   5.5 &  17.4 & 0.99 &  6400 &   65 &   22 &  1.28 & 4.16E-04 & 2.21E-02 &    300? &      a,b,c & ...	 &\\
   8122 &  10638 &       A3 &   6.7 &  71.7 & 1.57 &  8200 &   85 &   33 &  0.47 & 4.69E-04 &          &     100 &        a,d & ...	 &\\
   8241 &  10939 &      A1V &   5.0 &  57.0 & 1.94 & 10000 &   75 &   80 &  1.41 & 6.44E-05 & 4.52E-02 &    200? &        a,d & MIPS     &2,5\\
   9570 &  12471 &      A2V &   5.5 & 113.5 & 3.28 & 10000 &   85 &  105 &  0.93 & 1.01E-04 &          &     600 &        a,d & ...	 &\\
  10054 &  12467 &      A1V &   6.0 &  68.4 & 1.73 &  9200 &   60 &   94 &  1.38 & 8.72E-05 & 8.45E-02 &   200?? &        a,d & MIPS	 &2\\
  10670 &  14055 &    A1Vnn &   4.0 &  36.1 & 1.96 & 10000 &   75 &   80 &  2.24 & 7.18E-05 & 2.86E-02\tablenotemark{\ddag} &    100? &        a,d & ...	 &\\
  11360 &  15115 &       F2 &   6.8 &  44.8 & 1.23 &  7200 &   65 &   35 &  0.78 & 5.08E-04 & 4.48E-02\tablenotemark{\ddag} &    100? &      a,b,c & MIPS/\iso &\\
  11486 &  15257 &    F0III &   5.3 &  47.6 & 2.26 &  7400 &   85 &   39 &  0.84 & 1.14E-04 & 1.90E-02 & $\lesssim$1000 & a,d & ...	 &2\\
  11847 &  15745 &       F0 &   7.5 &  63.7 & 1.21 &  7600 &   85 &   22 &  0.35 & 1.72E-03 & 9.13E-02 &     30? &          d & MIPS/\iso &\\
  12361 &  16743 & F1III/IV &   6.8 &  60.0 & 1.58 &  7200 &   40 &  119 &  1.98 & 5.94E-04 &          &     200 &        a,d & MIPS     &\\
  12964 &  17390 &   F3IV/V &   6.5 &  45.1 & 1.39 &  7200 &   55 &   55 &  1.23 & 2.00E-04 & 8.52E-02 &   300?? &          a & MIPS	 &\\
  13005 &   ...  &       K0 &   8.1 &  67.7 & 2.17 &  5200 &   85 &   18 &  0.28 & 1.11E-03 &          &     ... &          b & ...	 &6,7\\
  13141 &  17848 &      A2V &   5.3 &  50.7 & 1.88 &  8200 &   55 &   97 &  1.92 & 6.39E-05 & 6.59E-02 &     100 &        a,d & MIPS     &\\
  14576 &  19356 &      B8V &   2.1 &  28.5 & 4.13 &  9200 &  250 &   13 &  0.46 & 1.67E-05 &          &    ...  &        ... & MIPS	 &2,7,8\\
  15197 &  20320 &      A5m &   4.8 &  36.8 & 2.00 &  7800 &   95 &   31 &  0.85 & 2.46E-05 & 2.59E-03 &    400? &        a,d & MIPS     &9\\
  16449 &  21997 &   A3IV/V &   6.4 &  73.8 & 1.57 &  9000 &   60 &   82 &  1.12 & 4.90E-04 & 2.24E-01\tablenotemark{\ddag} &     50? &        a,d & MIPS     &\\
  16537 &  22049 &      K2V &   3.7 &   3.2 & 0.69 &  5200 &   40 &   27 &  8.47 & 8.30E-05 & 2.61E-03\tablenotemark{\ddag} &     730 &      S2000 & MIPS/\iso &\\
  18437 &  24966 &      A0V &   6.9 & 103.5 & 1.50 & 10000 &   85 &   48 &  0.47 & 2.58E-04 &          &      10 &          d & ...	 &\\
  18859 &  25457 &      F5V &   5.4 &  19.2 & 1.19 &  6400 &   85 &   16 &  0.81 & 1.31E-04 & 3.68E-03 &      30 &      a,b,c & MIPS/\iso &\\
  18975 &  25570 &      F2V &   5.4 &  36.0 & 1.83 &  7000 &   85 &   28 &  0.80 & 8.86E-05 &          &     600 &     Hyades & ...	 &2,3\\
  19704 &  27346 &     A9IV &   7.0 & 114.5 & 2.57 &  7600 &   70 &   70 &  0.62 & 2.61E-04 &          &    600? &        a,d & ...	 &2,10,11\\
  19893 &  27290 &    F4III &   4.3 &  20.3 & 1.65 &  7200 &   80 &   31 &  1.53 & 2.30E-05 &          &    300? &        a,b & ...	 &12\\
  20635 &  27934 &   A7IV-V &   4.2 &  47.0 & 2.60 &  9000 &   85 &   67 &  1.44 & 4.72E-05 &          &     600 &     Hyades & ...	 &2,10\\
  21604 &  29365 &      B8V &   5.8 & 110.7 & 3.06 &  8800 &   75 &   97 &  0.88 & 3.78E-04 &          &    200? &        a,d & ...	 &2,8\\
  22226 &  30447 &      F3V &   7.9 &  78.1 & 1.31 &  7200 &   65 &   37 &  0.48 & 8.85E-04 & 1.33E-01 &  $\lesssim$100 &   a & MIPS     &\\
  22439 &  30743 &   F3/F5V &   6.3 &  35.4 & 1.46 &  6400 &   40 &   86 &  2.45 & 2.28E-04 &          & $>$1000 &      a,b,c & ...	 &2,13\\
  22845 &  31295 &      A0V &   4.6 &  37.0 & 1.67 &  9000 &   80 &   49 &  1.33 & 8.44E-05 & 2.22E-02 &    100? &        a,d & MIPS     &\\
  23451 &  32297 &       A0 &   8.1 & 112.1 & 1.24 &  8400 &   85 &   28 &  0.25 & 5.38E-03 & 4.62E-01 &     20? &          d & ...	 &\\
  24528 &  34324 &      A3V &   6.8 &  85.8 & 1.59 &  8800 &  100 &   28 &  0.33 & 1.72E-04 & 1.48E-02 &    200? &          d & ...	 &\\
  25197 &  34787 &     A0Vn &   5.2 & 104.3 & 3.26 & 10000 &  120 &   52 &  0.50 & 6.97E-05 & 2.07E-02 &    400? &        a,d & ...	 &2\\
  25790 &  36162 &     A3Vn &   5.9 & 105.6 & 2.92 &  8800 &   85 &   72 &  0.69 & 2.49E-04 &          &    600? &        a,d & ...	 &14\\
  26453 &  37484 &      F3V &   7.2 &  59.5 & 1.36 &  7000 &   90 &   19 &  0.32 & 2.85E-04 & 1.13E-02 &      30 &      a,b,c & MIPS     &\\
  26966 &  38206 &      A0V &   5.7 &  69.2 & 1.63 & 10000 &   85 &   53 &  0.76 & 1.99E-04 & 6.13E-02 &      50 &        a,d & MIPS     &\\
  27072 &  38393 &      F7V &   3.6 &   9.0 & 1.18 &  6600 &   90 &   15 &  1.64 & 7.71E-06 & 4.48E-04\tablenotemark{\ddag}  & $>$1000?? &      a,b & MIPS     &\\
  27288 &  38678 &   A2Vann &   3.5 &  21.5 & 1.65 &  9000 &  220 &    6 &  0.30 & 1.34E-04 &          &     100 &        a,d & MIPS     &\\
  27321 &  39060 &      A3V &   3.9 &  19.3 & 1.37 &  8600 &  110 &   19 &  1.01 & 2.64E-03 & 8.99E-02\tablenotemark{\ddag}  &      12 &$\beta$ Pic & MIPS     &\\
  27980 &  39833 &    G0III &   7.7 &  46.7 & 1.23 &  6000 &   70 &   20 &  0.45 & 2.79E-03 &          &     700 &      a,b,c & ...	 &2,15\\
  28103 &  40136 &      F1V &   3.7 &  15.0 & 1.52 &  7400 &  185 &    6 &  0.38 & 2.04E-05 &          &    300? &      a,b,d & MIPS     &\\
  28230 &  40540 &    A8IVm &   7.5 &  89.9 & 1.45 &  7800 &   90 &   25 &  0.28 & 6.06E-04 &          &     200 &          d & ...	 &2,16\\
  32480 &  48682 &      G0V &   5.2 &  16.5 & 1.08 &  6400 &   60 &   29 &  1.73 & 8.93E-05 &          &   600?? &        a,b & MIPS	 &17\\
  32775 &  50571 & F7III-IV &   6.1 &  33.2 & 1.38 &  6600 &   45 &   68 &  2.08 & 1.63E-04 & 8.26E-02 &    300? &      a,b,c & MIPS	 &\\
  33690 &  53143 &   K0IV-V &   6.8 &  18.4 & 0.88 &  5400 &   80 &    9 &  0.50 & 1.97E-04 & 1.87E-03 &    300? &      a,b,c & MIPS	 &\\
  34276 &  54341 &      A0V &   6.5 &  92.9 & 1.59 & 10000 &   85 &   51 &  0.55 & 2.01E-04 &          &      10 &          d & ...	 &18\\
  34819 &  55052 &     F5IV &   5.8 & 107.1 & 4.74 &  6800 &   45 &  251 &  2.35 & 1.01E-04 &          &   300?? &        a,c & ...	 &2,3\\
  35550 &  56986 &  F0IV... &   3.5 &  18.0 & 2.13 &  7200 &   60 &   71 &  3.95 & 8.93E-06 & 4.94E-03 &   400?? &      a,b,d & MIPS     &9\\
  36906 &  60234 &       G0 &   7.6 & 108.6 & 2.78 &  6200 &   85 &   34 &  0.32 & 4.29E-04 &          &    600? &        a,b & ...	 &2\\
  36948 &  61005 &   G3/G5V &   8.2 &  34.5 & 0.81 &  5600 &   60 &   16 &  0.48 & 2.58E-03 & 7.24E-02 &    100? &      a,b,c & MIPS	 &\\
  39757 &  67523 & F2mF5IIp &   2.8 &  19.2 & 3.41 &  6800 &   85 &   50 &  2.64 & 5.38E-06 &          &$\gtrsim$2000 & a,b,c & ...	 &\\
  40938 &  70298 &       F2 &   7.2 &  70.9 & 1.77 &  6800 &   85 &   26 &  0.37 & 3.54E-04 &          & $>$3000 &        a,b & ...	 &2\\
  41152 &  70313 &      A3V &   5.5 &  51.4 & 1.54 & 10000 &   80 &   56 &  1.09 & 5.24E-05 & 1.80E-02 &     200 &        a,d & MIPS     &\\
  41307 &  71155 &      A0V &   3.9 &  38.3 & 2.02 & 10000 &  130 &   29 &  0.73 & 4.09E-05 & 3.77E-03 &     100 &        a,d & MIPS     &\\
  42028 &  72660 &      A1V &   5.8 & 100.0 & 2.39 & 10000 &   85 &   77 &  0.77 & 7.07E-05 &          &     200 &        a,d & ...	 &2\\
  42430 &  73752 &   G3/G5V &   5.1 &  19.9 & 1.73 &  5800 &   80 &   21 &  1.06 & 3.21E-05 & 1.55E-03 &  $>$600 &      S2000 & ...	 &19\\
  43970 &  76543 &    A5III &   5.2 &  49.0 & 1.86 &  8800 &   85 &   46 &  0.94 & 1.04E-04 &          &    400? &        a,d & ...	 &2\\
  44001 &  76582 &     F0IV &   5.7 &  49.3 & 1.73 &  8000 &   85 &   35 &  0.72 & 2.22E-04 &          &   300?? &        a,d & ...	 &\\
  45758 &  80425 &       A5 &   6.6 &  98.1 & 2.43 &  7600 &   85 &   45 &  0.46 & 2.70E-04 &          &    300? &        a,d & ...	 &2\\
  48164 &  84870 &       A3 &   7.2 &  89.5 & 1.59 &  8000 &   85 &   32 &  0.37 & 5.48E-04 &          &     100 &          d & ...	 &1\\
  48541 &  85672 &       A0 &   7.6 &  93.1 & 1.19 &  9200 &   85 &   32 &  0.35 & 4.82E-04 &          &     30? &        a,d & ...	 &\\
  51438 &  91375 &    A2III &   4.7 &  79.4 & 3.10 & 10000 &   85 &   99 &  1.26 & 2.42E-05 &          &   400?? &        a,d & ...	 &20\\
  51658 &  91312 &     A7IV &   4.7 &  34.3 & 1.84 &  8200 &   40 &  179 &  5.23 & 1.06E-04 &          &     200 &        a,d & ...	 &9\\
  52462 &  92945 &      K1V &   7.7 &  21.6 & 0.77 &  5200 &   45 &   23 &  1.11 & 6.74E-04 & 3.91E-02 &     100 &        SBZ & MIPS     &\\
  53524 &  95086 &    A8III &   7.4 &  91.6 & 1.49 &  8200 &   85 &   32 &  0.35 & 1.49E-03 &          &      50 &          d & ...	 &2,21\\
  53910 &  95418 &      A1V &   2.3 &  24.3 & 2.84 & 10000 &  120 &   45 &  1.88 & 1.23E-05 & 2.73E-03 &     500 &        UMa & ...	 &\\
  53911 &   ...  &     K8Ve &  11.1 &  56.4 & 1.11 &  4000 &  140?&    2 &  0.04 & $>$2.17E-01 &          &       8 &      TWHya & MIPS     &\\
  55505 &  98800 &      K4V &   9.1 &  46.7 & 1.97 &  4200 &  160 &    3 &  0.07 & 1.12E-01 &          &       8 &        TWA & MIPS     &9\\
  56253 &  99945 &      A2m &   6.1 &  59.8 & 1.72 &  8200 &   85 &   37 &  0.62 & 1.04E-04 &          &    300? &        a,d & ...	 &\\
  56675 & 101132 &    F1III &   5.6 &  42.1 & 1.95 &  7000 &   50 &   88 &  2.11 & 1.42E-04 &          &     300 &    a,b,c,d & ...	 &2,22\\
  57632 & 102647 &   A3Vvar &   2.1 &  11.1 & 1.67 &  8800 &  160 &   11 &  1.06 & 4.25E-05 & 5.64E-04 &      50 &      S2001 & MIPS     &\\
  60074 & 107146 &      G2V &   7.0 &  28.5 & 0.97 &  6200 &   55 &   29 &  0.97 & 9.50E-04 & 8.99E-02\tablenotemark{\ddag}  &$\lesssim$100&  a,b,c & MIPS     &\\
  61174 & 109085 &      F2V &   4.3 &  18.2 & 1.62 &  6800 &  180 &    5 &  0.30 & 1.20E-04 &          &     300 &      a,b,c & MIPS     &\\
  61498 & 109573 &      A0V &   5.8 &  67.1 & 1.59 & 10000 &  110 &   30 &  0.46 & 4.43E-03 & 2.11E-01\tablenotemark{\ddag}  &       8 &   HR 4796A & MIPS     &\\
  61782 & 110058 &      A0V &   8.0 &  99.9 & 1.09 &  8800 &  130 &   11 &  0.12 & 2.54E-03 & 3.37E-02 &     10? &        LCC & IRS	 &21\\
  61960 & 110411 &      A0V &   4.9 &  36.9 & 1.49 &  9000 &   85 &   38 &  1.05 & 6.23E-05 & 9.86E-03 &   100?? &        a,d & MIPS     &\\
  63584 & 113337 &      F6V &   6.0 &  37.4 & 1.50 &  7200 &  100 &   18 &  0.48 & 1.01E-04 & 3.59E-03 &     50? &        a,b & ...	 &23\\
  64375 & 114576 &      A5V &   6.5 & 112.6 & 2.63 &  8200 &   85 &   56 &  0.51 & 3.90E-04 &          &     600 &        a,d & ...	 &1\\
  64921 & 115116 &      A7V &   7.1 &  85.4 & 1.53 &  8400 &   80 &   39 &  0.46 & 3.39E-04 &          &    100? &        a,d & ...	 &\\
  68101 & 121384 &      G8V &   6.0 &  38.1 & 2.95 &  5200 &   45 &   91 &  2.41 & 2.47E-04 &          & $>$3000 &      a,b,c & ...	 &\\
  68593 & 122652 &       F8 &   7.2 &  37.2 & 1.07 &  6400 &   60 &   28 &  0.76 & 1.36E-04 & 1.17E-02 &    300? &      a,b,c & MIPS     &\\
  69682 & 124718 &      G5V &   8.9 &  61.3 & 0.98 &  5800 &   85 &   10 &  0.17 & 2.11E-03 &          &  $>$500 &      a,b,c & ...	 &24\\
  69732 & 125162 &     A0sh &   4.2 &  29.8 & 1.72 &  9000 &  100 &   32 &  1.09 & 5.22E-05 & 5.86E-03 &    200? &        a,d & MIPS     &\\
  70090 & 125473 &     A0IV &   4.1 &  75.8 & 3.98 & 10000 &  120 &   64 &  0.85 & 2.11E-05 & 9.48E-03 &     300 &        a,d & ...	 &\\
  70344 & 126265 &    G2III &   7.2 &  70.1 & 2.12 &  6200 &   85 &   26 &  0.37 & 3.85E-04 &          &  $>$500 &        a,b & ...	 &\\
  70952 & 127821 &     F4IV &   6.1 &  31.7 & 1.30 &  6800 &   50 &   55 &  1.76 & 2.58E-04 & 8.26E-02\tablenotemark{\ddag} &    200? &        a,b & ...	 &\\
  71075 & 127762 & A7IIIvar &   3.0 &  26.1 & 3.08 &  8000 &   55 &  151 &  5.80 & 1.04E-05 &          &    1000 &        a,d & ...	 &\\
  71284 & 128167 &  F3Vwvar &   4.5 &  15.5 & 1.39 &  6600 &   40 &   88 &  5.70 & 4.91E-06 & 6.37E-03\tablenotemark{\ddag}  &  1000?? &      a,b,c & MIPS/\iso	 &\\
  73049 & 131625 &      A0V &   5.3 &  75.8 & 2.49 &  9000 &   85 &   64 &  0.86 & 7.39E-05 &          &     200 &        a,d & ...	 &2\\
  73145 & 131835 &     A2IV &   7.9 & 111.1 & 1.26 &  8600 &   90 &   26 &  0.24 & 3.07E-03 & 2.28E-01 &      10 &          d & ...	 &25\\
  73473 & 132742 &    B9.5V &   4.9 &  93.3 & 3.94 &  8800 &  150 &   31 &  0.34 & 7.22E-05 & 7.61E-03 &     500 &        a,d & ...	 &2,8,26\\
  73512 & 132950 &       K2 &   9.1 &  30.4 & 0.75 &  4800 &   85 &    5 &  0.18 & 1.17E-03 &          &  3000?? &        ... & ...	 &2\\
  74596 & 135502 &      A2V &   5.3 &  69.4 & 2.24 & 10000 &   65 &  123 &  1.77 & 3.26E-05 &          &     200 &        a,d & ...	 &\\
  74946 & 135382 &      A1V &   2.9 &  56.0 & 5.86 &  9400 &   50 &  481 &  8.60 & 9.29E-06 &          &   700?? &        a,d & ...	 &\\
  76127 & 138749 &    B6Vnn &   4.2 &  95.3 & 4.16 & 10000 &   75 &  171 &  1.80 & 1.99E-05 &          &    200? &        a,d & ...	 &\\
  76267 & 139006 &      A0V &   2.2 &  22.9 & 2.72 & 10000 &  190 &   17 &  0.76 & 2.41E-05 & 7.64E-04 &     500 &      a,b,d & MIPS     &8\\
  76375 & 139323 &      K3V &   7.6 &  22.3 & 0.85 &  5200 &   29 &   64 &  2.87 & 7.86E-04 &          &  5000?? &        a,b & ...	 &2,27\\
  76635 & 139590 &      G0V &   7.5 &  55.1 & 1.40 &  6200 &   85 &   17 &  0.31 & 3.93E-04 &          &  5000?? &        a,b & ...	 &\\
  76736 & 138965 &      A5V &   6.4 &  77.3 & 1.47 &  9600 &  140 &   16 &  0.21 & 1.17E-04 & 3.28E-03 &      20 &        a,d & MIPS     &2\\
  76829 & 139664 &   F5IV-V &   4.6 &  17.5 & 1.26 &  7000 &   75 &   25 &  1.46 & 1.15E-04 & 7.88E-03 &    200? &      a,b,c & MIPS     &\\
  77163 & 140775 &      A1V &   5.6 & 117.8 & 3.25 & 10000 &   40 &  472 &  4.01 & 1.39E-04 &          &     600 &        a,d & ...	 &28\\
  77542 & 141569 &       B9 &   7.1 &  99.0 & 1.49 &  9200 &  110 &   24 &  0.25 & 1.12E-02 & 3.32E-01\tablenotemark{\ddag}  &       5 &  HD 141569 & MIPS	 &\\
  78554 & 143894 &      A3V &   4.8 &  54.3 & 2.27 &  9000 &   45 &  211 &  3.89 & 4.64E-05 &          &     300 &        a,d & ...	 &\\
  81126 & 149630 &   B9Vvar &   4.2 &  92.7 & 4.91 &  9400 &   80 &  157 &  1.70 & 3.01E-05 &          &     700 &        a,d & ...	 &1\\
  81641 & 150378 &      A1V &   5.8 &  92.9 & 2.23 & 10000 &   95 &   57 &  0.62 & 1.24E-04 & 4.42E-02 &     200 &        a,d & ...	 &1\\
  81800 & 151044 &      F8V &   6.5 &  29.4 & 1.21 &  6200 &   55 &   35 &  1.22 & 8.30E-05 & 1.11E-02 &  $>$500 &        a,b & MIPS/\iso &\\
  82405 & 151900 & F1III-IV &   6.3 &  59.8 & 2.30 &  6600 &   85 &   32 &  0.54 & 2.98E-04 &          & $>$1000 &        a,d & ...	 &2,10\\
  83480 & 154145 &       A2 &   6.7 &  94.9 & 1.99 &  8400 &   85 &   45 &  0.48 & 4.28E-04 &          &    300? &          d & ...	 &28\\
  85157 & 157728 &     F0IV &   5.7 &  42.8 & 1.43 &  8600 &   90 &   30 &  0.71 & 2.67E-04 & 2.63E-02 &    100? &        a,d & ...	 &\\
  85537 & 158352 &      A8V &   5.4 &  63.1 & 2.52 &  8400 &   70 &   85 &  1.35 & 6.81E-05 & 5.39E-02 &    600? &        a,d & MIPS	 &\\
  87108 & 161868 &      A0V &   3.7 &  29.1 & 1.91 &  9400 &   85 &   54 &  1.87 & 7.84E-05 & 2.51E-02 &    200? &        a,d & MIPS     &\\
  87558 & 162917 &   F4IV-V &   5.8 &  31.4 & 1.50 &  6600 &   85 &   20 &  0.67 & 2.49E-04 &          &    400? &      a,b,c & ...	 &\\
  88399 & 164249 &      F5V &   7.0 &  46.9 & 1.27 &  6800 &   70 &   27 &  0.60 & 1.03E-03 & 8.23E-02 &      12 &       ZSBW & MIPS/\iso &\\
  90185 & 169022 &  B9.5III &   1.8 &  44.3 & 6.66 & 10000 &  100 &  155 &  3.50 & 4.46E-06 &          &   300?? &      a,b,d & ...	 &\\
  90936 & 170773 &      F5V &   6.2 &  36.1 & 1.34 &  7000 &   50 &   61 &  1.69 & 4.63E-04 & 1.89E-01 &    200? &      a,b,c & MIPS/\iso	 &\\
  91262 & 172167 &   A0Vvar &   0.0 &   7.8 & 2.58 & 10000 &   80 &   93 & 12.10 & 2.14E-05 & 8.37E-03\tablenotemark{\ddag}  &     220 &       Vega & MIPS	 &\\
  92024 & 172555 &      A7V &   4.8 &  29.2 & 1.52 &  8000 &  320 &    2 &  0.08 & 8.10E-04 &          &      12 &       ZSBW & MIPS     &\\
  93542 & 176638 &     A0Vn &   4.7 &  56.3 & 2.11 & 10000 &  120 &   34 &  0.60 & 9.70E-05 & 1.23E-02 &    200? &        a,d & ...	 &\\
  95261 & 181296 &     A0Vn &   5.0 &  47.7 & 1.61 &  9600 &  150 &   15 &  0.32 & 2.13E-04 & 5.25E-03 &      12 &       ZSBW & MIPS/\iso &\\
  95270 & 181327 &   F5/F6V &   7.0 &  50.6 & 1.39 &  6600 &   75 &   25 &  0.50 & 3.47E-03 & 2.38E-01 &      12 &       ZSBW & MIPS     &\\
  95619 & 182681 &   B8/B9V &   5.7 &  69.1 & 1.71 & 10000 &   85 &   55 &  0.80 & 1.95E-04 &          &     50? &        a,d & ...	 &\\
  96468 & 184930 &    B5III &   4.3 &  94.3 & 4.01 & 10000 &   60 &  259 &  2.75 & 3.52E-05 &          &     ... &        ... & ...	 &2,7\\
  99273 & 191089 &      F5V &   7.2 &  53.5 & 1.39 &  6600 &   95 &   15 &  0.29 & 1.39E-03 & 3.43E-02 &$\lesssim$30&  a,b,c  & MIPS     &9\\
  99473 & 191692 &  B9.5III &   3.2 &  88.0 & 6.63 & 10000 &   85 &  213 &  2.43 & 6.60E-06 &          &    500? &        a,d & ...	 &\\
 101612 & 195627 &    F1III &   4.8 &  27.6 & 1.70 &  7400 &   65 &   51 &  1.86 & 1.11E-04 & 3.17E-02 &    200? &        a,d & MIPS     &\\
 101769 & 196524 &     F5IV &   3.6 &  29.9 & 3.63 &  6800 &  130 &   23 &  0.77 & 1.56E-05 & 9.05E-04 &   200?? &      a,b,c & ...	 &2,9\\
 101800 & 196544 &      A2V &   5.4 &  54.3 & 1.65 &  9000 &  100 &   31 &  0.57 & 3.86E-05 & 4.07E-03 &      30 &        a,d & MIPS     &9\\
 102409 & 197481 &     M1Ve &   8.8 &   9.9 & 0.86 &  3500 &   50 &    9 &  0.98 & 3.64E-04 & 8.80E-03\tablenotemark{\ddag}  &      12 &       ZSBW & MIPS     &\\
 103752 & 199475 &      A2V &   6.4 &  83.3 & 1.83 &  8800 &   85 &   45 &  0.55 & 2.45E-04 &          &     200 &        a,d & ...	 &2\\
 105570 & 203562 &      A3V &   5.2 & 110.4 & 4.02 &  9000 &   85 &  104 &  0.95 & 8.80E-05 &          &    600? &        a,d & ...	 &1\\
 106741 & 205674 &  F3/F5IV &   7.2 &  52.6 & 1.22 &  7200 &   85 &   20 &  0.39 & 3.96E-04 &          &    300? &        a,b & ...	 &\\
 107022 & 205536 &      G8V &   7.1 &  22.1 & 0.89 &  5600 &   80 &   10 &  0.46 & 2.92E-04 & 3.20E-03 &  $>$500 &        a,b & ...	 &\\
 107412 & 206893 &      F5V &   6.7 &  38.9 & 1.24 &  6600 &   55 &   41 &  1.07 & 2.72E-04 & 3.18E-02\tablenotemark{\ddag} &    200? &        a,b & MIPS/\iso &\\
 107649 & 207129 &      G2V &   5.6 &  15.6 & 0.98 &  6000 &   55 &   27 &  1.74 & 1.21E-04 & 9.67E-03 &     600 &        SZB & MIPS/\iso	 &\\
 108809 & 209253 &   F6/F7V &   6.6 &  30.1 & 1.10 &  6200 &   75 &   18 &  0.58 & 7.33E-05 & 2.60E-03 &   200?? &      a,b,c & MIPS/\iso	 &\\
 109857 & 211336 &     F0IV &   4.2 &  25.7 & 1.86 &  7800 &   65 &   62 &  2.41 & 1.56E-04 & 6.58E-02 &    600? &      a,c,d & ...	 &\\
 110867 & 210681 &    K0III &   8.1 &  61.8 & 1.87 &  5200 &   85 &   16 &  0.26 & 7.15E-04 &          &     ... &        ... & ...	 &2,7\\
 111278 & 213617 &      F1V &   6.4 &  52.9 & 1.57 &  7600 &   55 &   69 &  1.32 & 9.35E-05 & 4.9E-02  &    600? &        a,d & MIPS     &\\
 113368 & 216956 &      A3V &   1.2 &   7.7 & 1.81 &  8600 &   65 &   73 &  9.60 & 7.98E-05 & 2.41E-02\tablenotemark{\ddag}  &     220 &  Fomalhaut & MIPS	 &\\
 114189 & 218396 &      A5V &   6.0 &  39.9 & 1.37 &  7800 &   50 &   77 &  1.94 & 2.29E-04 & 1.00E-01\tablenotemark{\ddag} &      30 &        a,d & \iso	 &\\
 116431 & 221853 &       F0 &   7.3 &  71.2 & 1.48 &  7400 &   85 &   26 &  0.37 & 7.38E-04 & 5.47E-02 &$\lesssim$100 &     a & MIPS/\iso &\\
\enddata
\tablecomments{Calculations use 1\,AU=215\,$R_\odot$}
\tablenotetext{\dag}{Age Methods: S2000: \citet{son00}; S2001: \citet{son01};
SBZ: \citet{son02a}; ZFK: \citet{zuc95b}; ZSBW: \citet{zsbw01}; 
ZSW: \citet{zsw01}; ZW: \citet{zw00}; SZB: \citet{szb03}; 
a: \textit{UVW} ; \citep{zs04b}; b: X-ray emission; e.g., \citet{szb03};
c: lithium age; \citep{szb03}; 
d: location on an A-star Hertzsprung-Russell diagram; \citep{low00}}
\tablenotetext{\ddag}{Dust mass measurements are directly from submillimeter observations.}
\tablenotetext{\alpha}{
1. binary.\\
2. New debris disk candidate.\\
3. Caution: \iras SCANPI shows high background fluctuation near \iras 60\,$\micron$ detection.\\
4. HIP~7345 (=49~Cet) is the only known main-sequence A-type star with CO emission 
   detected with a radio telescope (ZFK), thus suggesting a very young age. But its 
   galactic space motion \textit{UVW} ($-23,-17,-4$) with respect to the Sun is not indicative 
   of extreme youth (U is positive toward the Galactic center). \\
5. HIP 8241 shows the age of the Pleiades on an A star HR diagram \citep{low00} but of the 
   Hyades in \textit{UVW} measurements.\\
6. There is a galaxy at $\sim$48\,$\arcsec$ East of HIP 13005 in the cross-scan direction as described 
   in Paper I.  However, a more careful check of the \iras 60\,$\micron$ offset using the FSC 
   long format indicates that both \iras 12\,$\micron$ and 60\,$\micron$ detections have the 
   same offsets away from the galaxy in the same cross-scan direction.  Thus, we include 
   HIP~13005 with a caution. \\
7. No age estimate is given for HIP~13005, HIP~14576, HIP~96468, \& HIP~110867 \\
8. Eclipsing binary of the Algol type.\\
9. Spectroscopic binary.\\
10. Caution: \iras SCANPI shows $>$ 30\,$\arcsec$ offset \iras 60\,$\micron$ detection from 
    the stellar position in in-scan direction. \\
11. There are two FSC detections for HIP~19704 separated by 34\,$\arcsec$ in the in-scan direction.  
    One has 12 and 25\,$\micron$ detections, the other has a 60\,$\micron$ detection.  The long 
    format of FSC locates the 60\,$\micron$ source on HIP 19704.\\
12. In addition to the point-like 60\,$\micron$ source reported in the FSC, there is 
    an extended optical source 70\,$\arcsec$ from the \iras position of HIP 19893 in 
    the in-scan direction.  \citet{jur04} found no strong excess up to 35\,$\micron$
    in this star.  Thus the \iras excess at 60\,$\micron$ should be regarded with caution.\\
13. Caution: there is a galaxy 90\,$\arcsec$ East of HIP 22439.\\
14. Caution: there is a galaxy 55\,$\arcsec$ East of the FSC position
    at the 3\,$\sigma$ edge of the error ellipse, mostly in the cross-scan direction. \\ 
15. Caution: \iras FSC detection is 40\,$\arcsec$ West of HIP 27980, and \iras SCANPI profile 
    is very broad.\\
16. Caution: there is a galaxy 58\,$\arcsec$ away from the \iras position of HIP 28230 in the cross-scan 
    direction.\\
17. There is a ROSAT All-Sky Survey X-ray source $\sim$44\,$\arcsec$ from HIP~32480, but 
    \textit{UVW} indicates an old age. \\
18. Location on A-star HR diagram near HR 4796 is suggestive of a 10 Myr age, 
    but the V component of UVW (-16,-44,-9;\citealt{moo06}) is quite unlike 
    that of most very young stars. \\
19. HIP~42430 is a $1\farcs0$ binary.\\
20. Caution: \iras SCANPI shows a bad profile fit to the 60\,$\micron$ source.\\
21. \citet{kou05} say HIP~53524 and HIP~61782 are LCC members.\\
22. Caution: \iras SCANPI shows no source detection.\\
23. The M-star companion LDS\,2662 to HIP~63584 is very young based on its location 
    on an $M_K$ versus $V-K$ color magnitude diagram (e.g., Figure~2 in SZB).\\
24. \citet{moo06} rejected HIP 69682 based on a nearby 2MASS source with an excess
    in the $K_{s}$-band.  However, no NED identified extended source exists within 
    2\,$\arcmin$ from this star and the FSC long format indicates that the 60\,$\micron$ 
    detection falls on the star itself.  
    The Galactic space motion (\textit{UVW}) and absence of lithium and of X-ray emission 
    all point to an old star. There is no evidence on the Digital Sky Survey and 
    2MASS All Sky QuickLook Images ($JHK_s$) of a nearby galaxy. Yet $\tau$ is 
    very large. \\
25. HIP 73145 is an Upper Centaurus Lupus member.\\
26. HIP~73473 has significant X-ray flux.\\
27. Caution: there exists a large galaxy at $\sim$80\,$\arcsec$ East of HIP 76375.\\
28. \citet{moo06} rejected HIP 77163 \& HIP 83480 based on their location near 
    the wall of the Local Bubble.\\
}
\end{deluxetable}

\clearpage
\begin{deluxetable}{ccccccc}
\tablenum{3}
\tablecolumns{7}
\tablecaption{Algols from \iras and \spit \label{tbl-3}}
\tablewidth{0pc}
\tablehead{
\colhead{HIP}                &\colhead{HD}               &\colhead{Other}      &
\multicolumn{2}{c}{Rieke et al.\ (2005)}                 &\colhead{This Paper} &
\colhead{Triple} \\
\colhead{}                   &\colhead{}                 &\colhead{}           &
\colhead{Observed?}          &\colhead{Excess?}          &\colhead{Excess?}    &
\colhead{System?}        
}
\startdata
14576 &	19356  & Algol A      & yes & 1.07 (no)	& marginal?	& yes \\
21604 &	29365  & HU Tau	      & no  &	        & strong	& ?   \\
28360 &	40183  & $\beta$ Aur  & yes & 0.88 (no)	& nothing	& ?   \\
73473 &	132742 & $\delta$ Lib & no  &	        & 2 wavelengths	& yes \\
76267 &	139006 & $\alpha$ CrB & yes & 1.29 (yes)& 2 wavelengths & ?   \\
\enddata
\end{deluxetable}

\end{document}